\documentclass[11pt]{article}
\usepackage{amsmath}
\usepackage{amssymb}
\usepackage{amsfonts}
\usepackage{graphicx}
\usepackage{subcaption}
\usepackage{fullpage}
\usepackage{cleveref}
\usepackage{cellspace}
\usepackage{scalerel}
\usepackage{stackengine}

\newcommand{\xrog}{\mathring{x}}
\newcommand{\xrogvec}{\mathring{\bold{x}}}
\newcommand{\rhorog}{\mathring{\rho}}
\newcommand{\urogf}{\mathring{u}''}
\newcommand{\Prog}{\mathring{P}}

\newcommand{\Trog}{\mathring{T}}
\newcommand{\lbra}{\left <}
\newcommand{\rbra}{\right >}

\newcommand{\ka}{\stackon[-8pt]{$u''^{(\alpha)}_i u''^{(\alpha)}_i$}{\vstretch{1.5}{\hstretch{2.0}{\widetilde{\phantom{\;\;\;\;\;\;\;\;}}}}}}
\newcommand{\ks}{\stackon[-8pt]{$u''^{(s)}_i u''^{(s)}_i$}{\vstretch{1.5}{\hstretch{2.0}{\widetilde{\phantom{\;\;\;\;\;\;\;\;}}}}}}
\newcommand{\kd}{\stackon[-8pt]{$u''^{(d)}_i u''^{(d)}_i$}{\vstretch{1.5}{\hstretch{2.0}{\widetilde{\phantom{\;\;\;\;\;\;\;\;}}}}}}

\title{Self-consistent feedback mechanism for the sudden viscous dissipation of finite-Mach-number compressing turbulence}
\author{A. Campos, B. Morgan}
\date{}

\begin{document}

\maketitle

Previous work (S.\@ Davidovits and N.\@ J.\@ Fisch, ``Sudden viscous dissipation of compressing turbulence,'' Phys.\@ Rev.\@ Lett., 116(105004), 2016) demonstrated that the compression of a turbulent field can lead to a sudden viscous dissipation of turbulent kinetic energy (TKE), and suggested this mechanism could potentially be used to design new fast-ignition schemes for inertial confinement fusion. We expand on previous work by accounting for finite Mach numbers, rather than relying on a zero-Mach-limit assumption as previously done. The finite-Mach-number formulation is necessary to capture a self-consistent feedback mechanism in which dissipated TKE increases the temperature of the system, which in turn modifies the viscosity and thus the TKE dissipation itself. Direct numerical simulations with a tenth-order accurate Pad\'e scheme were carried out to analyze this self-consistent feedback loop for compressing turbulence. Results show that, for finite Mach numbers, the sudden viscous dissipation of TKE still occurs, both for the solenoidal and dilatational turbulent fields. As the domain is compressed, oscillations in dilatational TKE are encountered due to the highly-oscillatory nature of the pressure dilatation. An analysis of the source terms for the internal energy shows that the mechanical work term dominates the viscous turbulent dissipation. As a result, the effect of the suddenly dissipated TKE on temperature is minimal for the Mach numbers tested. Moreover, an analytical expression is derived that confirms the dissipated TKE does not significantly alter the temperature evolution for low Mach numbers, regardless of compression speed. The self-consistent feedback mechanism is thus quite weak for subsonic turbulence, which could limit its applicability for inertial confinement fusion.

\section{Introduction}
The compression of a turbulent flow occurs in a broad array of applications. Examples include one-dimensional compressions in internal combustion engines \cite{han1995} or across shock waves \cite{larsson2013}, axisymmetric compressions in Z-pinches \cite{kroupp2018}, spherically-symmetric compressions in inertial confinement fusion \cite{weber2014,haines2014}, and three-dimensional complex contractions in the interstellar medium \cite{robertson2012}. Moreover, the compression mechanism often leads to complex turbulence dynamics, and the resulting evolution of turbulence can have a strong effect on the overall behavior of the application under consideration. Thus, increased levels of understanding and improved modeling capabilities for this phenomenon are essential. 

Numerous direct numerical simulations of compressing turbulence have been previously carried out with the aim of improving engineering turbulence models; see for example \cite{wu1985, coleman1991,blaisdell1991,coleman1993,blaisdell1996}. These studies treated the fluid as a traditional gas, for which the dependence of viscosity $\mu$ on temperature $T$ is given by $\mu \sim T^n$, with $n$ having a value of, or close to, $3/4$. On the other hand, \cite{davidovits2016} demonstrated, through computational simulations, that when a power law exponent characteristic of weakly-coupled plasmas is used, i.e.\@ $n=5/2$, a sudden viscous mechanism occurs which dissipates the turbulent energy. Their results showed that a turbulent field subjected to a continuous isotropic compression initially creates an amplification of turbulent kinetic energy (TKE), until viscous scaling dominates and TKE is rapidly dissipated into heat. It was thus proposed in \cite{davidovits2016} that the resulting increases of temperature could be used to improve the ignition conditions for inertial confinement fusion. 

Subsequent work has expanded on the simulations of \cite{davidovits2016}. The effect of ionization on the scaling of viscosity was accounted for in \cite{davidovits2016b}. For that study, the ionization state $Z$ was assumed to depend solely on temperature, and thus the plasma viscosity $\mu \sim T^{5/2} / Z^4$ was simplified to the form $\mu \sim T^\beta$. Their analysis of the evolution of the energy spectrum showed that the sudden dissipation of TKE occurs for $\beta > 1$ only. A TKE model that accounts for the viscous dissipative mechanism for isotropic compressions is presented in \cite{davidovits2017}. This model was validated against direct numerical simulations, and showed excellent agreement for viscosity-power-law exponents greater than one. The model was then used to estimate the partition of energy between the turbulence and heat, as the compression proceeds in time. A two-point spectral model based on the EDQNM formulation was used by \cite{viciconte2018}, along with direct numerical simulations, to reproduce the sudden viscous dissipation mechanism. The lower computational cost of the EDQNM model allowed for the analysis of high-Reynolds-number effects and thus the identification of three distinct regimes: turbulent production, non-linear energy transfer, and viscous dissipation. Moreover, the assumption of homogeneous turbulence was relaxed and a spherical inhomogeneous turbulent layer under compression was simulated with both DNS and EDQNM closures. The sudden dissipation of TKE was also observed for this new case. Finally, in \cite{davidovits2018}, a stability boundary for hot spot turbulence was derived to demarcate states of the compression for which a decrease of TKE is guaranteed. Moreover, an upper limit for the amount of TKE that can be generated during a compression was proposed. This upper limit was then compared to the internal thermal energy of the system. 

The simulations of the sudden viscous dissipation mechanism previously conducted have relied on the zero-Mach-limit assumption. Given a decomposition of the velocity $U_i = \lbra U_i \rbra + u_i$, where $\lbra U_i \rbra$ is the Reynolds-averaged mean flow and $u_i$ the fluctuating velocity, the governing equations for the fluctuations in the zero-Mach limit take the form \cite{wu1985}
 \begin{equation}
 \label{eq:lowmach_1}
 \frac{\partial u_i}{\partial x_i} = 0,
 \end{equation}
 \begin{equation}
 \label{eq:lowmach_2}
\lbra \rho \rbra \left ( \frac{\partial u_i}{\partial t} + u_j \frac{\partial u_i}{\partial x_j} \right)= - \frac{\partial P}{\partial x_i} + \mu \frac{\partial^2 u_i}{\partial x_j \partial x_j} + f_i.
\end{equation}
In the above, $\lbra \rho \rbra$ is the Reynolds-averaged density, $P$ the pressure, $\mu$ the viscosity, and $f_i$ a forcing function that accounts for the effect of the compression. The viscosity depends on temperature and thus an a-priori time evolution for temperature needs to be provided. For an adiabatic isotropic compression, this is
\begin{equation}
\label{eq:adiabatic_temp}
T = T_0 L^{-2},
\end{equation}
where $L$ is a characteristic length of the domain being compressed and $T_0$ the initial temperature. 

The approach described above is suitable for demonstrating the sudden dissipation of TKE, but does not capture the self-consistent feedback mechanism mentioned in \cite{davidovits2016}. The feedback mechanism begins with an energy transfer from the TKE towards the internal energy, as a result of the sudden viscous dissipation. This, in turn, causes increased temperatures that amplify the viscosity of the system. The stronger values of viscosity then precipitate the viscous dissipation of TKE, thus completing the self-consistent feedback loop. In the zero-Mach limit, an evolution equation for the internal energy is not solved, and thus the effect of the dissipated TKE on the internal energy and the viscosity cannot be reproduced. It is expected that accounting for the feedback mechanism would lead to viscous dissipations that are more sudden, or of increased intensity \cite{davidovits2016}. An alternative to the assumption of the zero-Mach limit is turbulence belonging to the finite-Mach number regime. For this case, fully coupled governing equations for density, velocity, and energy are solved, which allows for an explicit accounting of the forward transfer of dissipated TKE into heat, and the subsequent effect of increased temperature and viscosity on the dissipation. The focus of this study is the simulation of turbulence in the finite-Mach number regime to investigate the complex self-consistent feedback mechanism, and thus further assess the benefits of viscous dissipation for inertial-confinement fusion and other high-energy density applications.

The outline of the paper is as follows. \Cref{sec:governing_equations} includes a description of the governing equations for turbulence in the finite-Mach number regime. The mechanisms that account for the energy transfer between the TKE and the internal energy are also discussed. In \cref{sec:comp_details}, details of the numerical simulations, such as the discretization scheme and the creation of realistic initial conditions, are included. The results of the simulations are then provided in \cref{sec:results}, which is divided into two subsections. \Cref{sec:tke_results} focuses on the component of the feedback mechanism related to the TKE. Thus, the evolution of the TKE, its budget, and spectra are analyzed in this subsection. The component of the feedback loop associated with the internal energy is then investigated in \cref{sec:ie_results}, where an analysis of the temperature evolution and sources for the internal energy are included. Finally, the paper ends with \cref{sec:conclusion}, where concluding remarks and a discussion of future work is provided.

\section{Governing equations}
\label{sec:governing_equations}
\subsection{Navier-Stokes equations for isotropic compressions}
We denote $\widetilde{U}_i$ and $u''_i$ as the Favre-averaged and Favre-fluctuating velocities, respectively, so that $U_i = \widetilde{U}_i + u''_i$ \cite{wilcox2010}. In analogy to the zero-Mach-limit formulation of \cite{davidovits2016}, we analyze the effect of a compression on a statistically homogeneous turbulent field $u''_i$, where the compression is achieved through a specified Favre-averaged mean flow $\widetilde{U}_i$. The Favre-averaged velocity for homogeneous compressible turbulence needs to be restricted to the form $\widetilde{U}_i = G_{ij} x_j$ \cite{blaisdell1991}. The deformation tensor $G_{ij}$ corresponding to an isotropic compression is given in \cite{wu1985, blaisdell1991, davidovits2016}, and can be expressed as 
\begin{equation}
\label{eq:def_tensor}
G_{ij} = \frac{\dot{L}}{L} \delta_{ij},
\end{equation}
where $L$ is a time-dependent characteristic length of the compressed domain, and $\dot{L}$ is the constant time rate of change of $L$. Given this formalism, one can derive, as detailed in \cref{sec:comp_fluc_NS_mean_comp}, a set of Navier-Stokes equations for the fluctuating velocity undergoing mean-flow compression. These equations, which are summarized below, constitute the finite-Mach-number analog of the low-Mach-number \cref{eq:lowmach_1,eq:lowmach_2,eq:adiabatic_temp}, 
\begin{equation}
\label{eq:rho_miranda}
\frac{\partial \rho}{\partial t} + \frac{\partial \rho u''_i}{\partial x_i} = f^{(\rho)},
\end{equation}
\begin{equation}
\label{eq:rhou_miranda}
\frac{\partial \rho u''_i}{\partial t} + \frac{\partial \rho u''_i u''_j}{\partial x_j} = \frac{\partial \sigma_{ij}}{\partial x_j} + f_i^{(u)},
\end{equation}
\begin{equation}
\label{eq:rhoE_miranda}
\frac{\partial \rho E_t}{\partial t} + \frac{\partial \rho E_t u''_i}{\partial x_i} = \frac{\partial u''_i \sigma_{ij}}{\partial x_j} + \frac{\partial}{\partial x_j} \left ( \kappa \frac{ \partial T}{\partial x_j} \right ) + f^{(e)}.
\end{equation}
In the above $\rho$ is the density, $u''_i$ the Favre-fluctuating velocity, and $T$ the temperature. $E_t$ is the total energy, which is given by $E_t = U + K$, where $U=C_vT$ is the internal energy and $K_t = \frac{1}{2} u''_i u''_i$ is the kinetic energy associated with the turbulent fluctuations. $C_v$ is the specific heat at constant volume. The stress tensor is given by
\begin{equation}
\sigma_{ij} = -P \delta_{ij} + 2 \mu \left [ \frac{1}{2} \left ( \frac{\partial u''_i}{\partial x_j} + \frac{\partial u''_j}{\partial x_i} \right ) - \frac{1}{3} \frac{\partial u''_k}{\partial x_k} \delta_{ij} \right ] ,
\end{equation}
where $P$ is the pressure and $\mu$ the viscosity. A power law of the form $\mu = \mu_0 \left (T / T_0\right)^n$ is used, where $\mu_0$ and $T_0$ represent reference viscosity and temperature values, and $n$ is the power-law exponent. The thermal conductivity $\kappa$ is computed according to $\kappa = \mu C_p / Pr$, where $C_p$ is the specific heat at constant pressure and $Pr$ the Prandtl number. An ideal equation of state $P = \rho R T$ is used, where $R$ is the ideal gas constant. The forcing functions $f^{(\rho)}$, $f^{(u)}_i$, and $f^{(e)}$ account for the effect of the mean compression on the density, velocity, and total energy, respectively, and are defined as
\begin{equation}
\label{eq:frho_miranda}
f^{(\rho)} = -2 \dot{L} \rho,
\end{equation}
\begin{equation}
\label{eq:frhou_miranda}
f^{(u)}_i = -3 \dot{L} \rho u''_i,
\end{equation}
\begin{equation}
\label{eq:fe_miranda}
f^{(e)} = - \left [2 \rho E_t + \rho u''_i u''_i + 3 P \right ] \dot{L} .
\end{equation}
The equations above are suitable for numerical simulations, since the compressive effect of the mean flow $\widetilde{U}_i$ has been abstracted into the three forcing functions above. These are the equations solved for the current study.

\subsection{Energy exchange for compressible turbulence}
For turbulence in the finite-Mach-number regime, the Helmholtz decomposition is often employed to express the fluctuating velocity as $u''_i = u''^{(s)}_i + u''^{(d)}_i$, where $u''^{(s)}_i$ and $u''^{(d)}_i$ are the solenoidal and dilatational velocities, respectively. The solenoidal component satisfies $\nabla \times \bold{u}''^{(s)} = \bold{w}$ and $\nabla \cdot \bold{u}''^{(s)} = 0$, where $\bold{w} =\nabla \times \bold{u}''$ is the vorticity vector, and the dilatational component satisfies $\nabla \times \bold{u}''^{(d)} = 0$ and $\nabla \cdot \bold{u}''^{(d)} = d$, where $d = \nabla \cdot \bold{u}''$ is the dilatation. 

Given this decomposition, two TKEs can be defined. These are the solenoidal TKE
\begin{equation}
k^{(s)} = \frac{1}{2} \ks,
\end{equation}
the dilatational TKE
\begin{equation}
k^{(d)} = \frac{1}{2} \kd.
\end{equation}
There are two additional energies in the system, namely, the mean kinetic energy
\begin{equation}
\bar{K} = \frac{1}{2} \widetilde{U}_i \widetilde{U}_i,
\end{equation}
and the mean internal energy
\begin{equation}
\widetilde{U} = C_v \widetilde{T}.
\end{equation}

\begin{table}
\centering
\caption{Sources in the evolution equations for the solenoidal, dilatational, mean, and internal energies. Upper scripts $\alpha$ stand for either $s$ or $d$. $\tau_{ij}$ represents the Favre-averaged Reynolds stresses ($\tau_{ij} = \widetilde{u''_i u''_j}$).}
\label{tb:energy_sources}
\begin{tabular}{ Sc Sc Sc }
Name & Symbol & Definition \\
\hline\hline
Intermode advection & $IA^{(\alpha)}$ & $- \left < \frac{\partial \sqrt{\rho} u''_i u''_j}{\partial x_j} \sqrt{\rho} u''^{(\alpha)}_i \right > + \left < \frac{\rho u''_i u''^{(\alpha)}_i}{2} d \right >$ \\
Production & $P^{(\alpha)}$ & $- \frac{2}{3} \lbra \rho \rbra k^{(\alpha)} G_{ii}$ \\
Solenoidal dissipation & $\lbra \rho \rbra \epsilon^{(s)}$ & $\left < \mu w_i w_i \right >$  \\
Dilatational dissipation & $\lbra \rho \rbra \epsilon^{(d)}$ & $\frac{4}{3} \left < \mu d^2 \right >$ \\
Pressure dilatation & $PD$ & $\left < P d \right >$ \\
Mean kinetic energy advection & $AD^{(\bar{K})}$ & $\lbra \rho \rbra \widetilde{U}_j \frac{ \partial \bar{K}}{\partial x_j}$ \\
Mean kinetic energy transport & $T^{(\bar{K})}$ & $\frac{\partial}{\partial x_j} \left ( \widetilde{U}_i \lbra \rho \rbra \tau_{ij} + \widetilde{U}_j \lbra P \rbra \right )$ \\
Mechanical work & $MW$ & $-\lbra P \rbra G_{ii}$\\
\end{tabular}
\end{table}

\begin{figure}[!t]
    \centering
    \includegraphics[width=\textwidth]{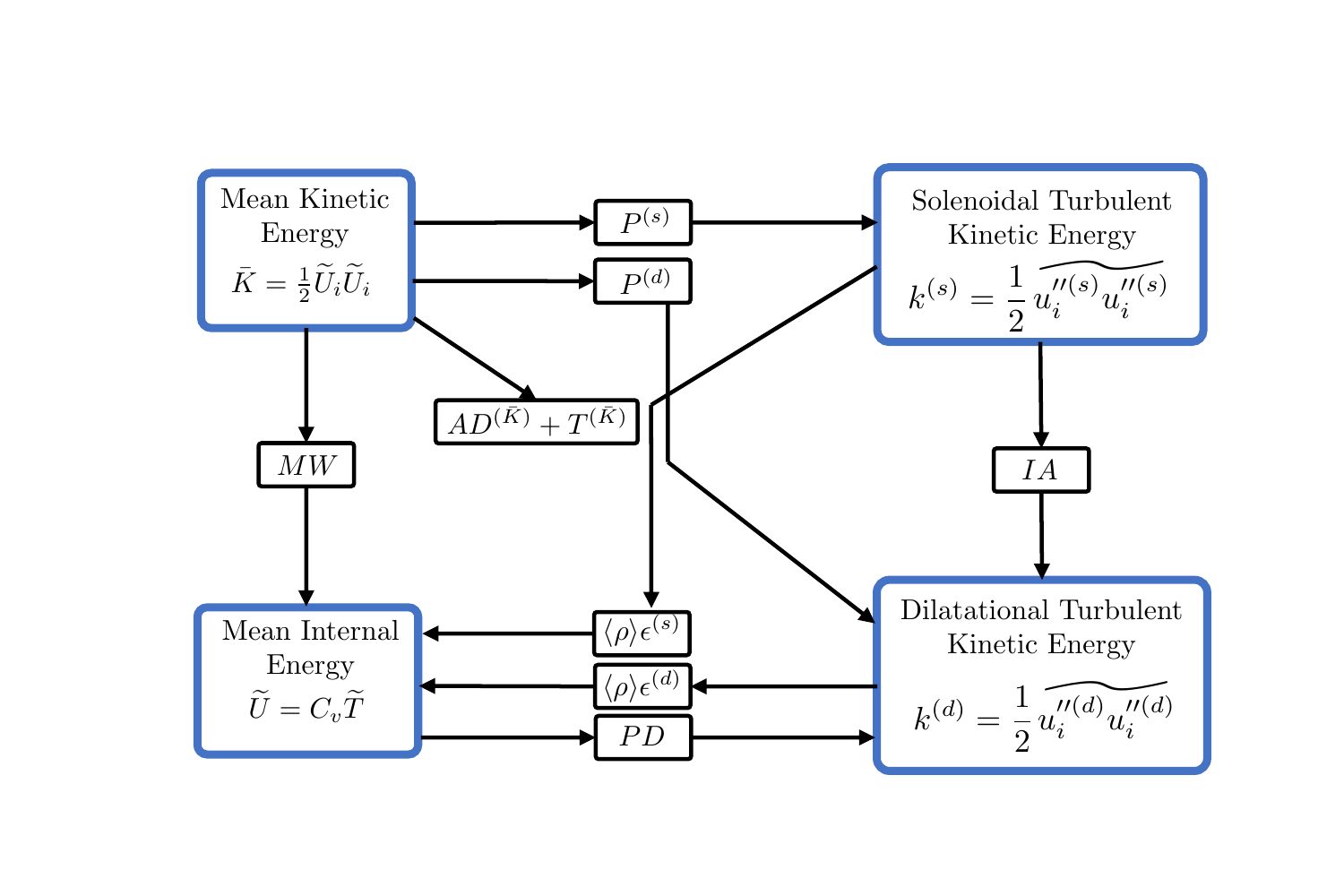}
    \caption{Schematic of energy transfer between the four energy components: mean kinetic energy, mean internal energy, solenoidal TKE, and dilatational TKE.}
    \label{fig:energy_chart}
\end{figure}

A derivation of the governing equations for the solenoidal and dilatational TKEs is given in \cref{sec:sol_dil_evolution}. Along with the evolution equations for $\bar{K}$ and $\widetilde{U}$, one can summarize the governing dynamics of the four relevant energies for homogeneous turbulence as follows
\begin{align}
\lbra \rho \rbra \frac{d k^{(s)}}{d t} &= IA^{(s)} +P^{(s)} - \langle \rho \rangle \epsilon^{(s)},  \label{eq:ks_evolution}\\
\lbra \rho \rbra \frac{d k^{(d)}}{d t} &= IA^{(d)} +P^{(d)} - \langle \rho \rangle \epsilon^{(d)} + PD, \label{eq:kd_evolution}\\
\lbra \rho \rbra \frac{\partial \bar{K} }{\partial t} &= -AD^{(\bar{K})} - T^{(\bar{K})} - MW - P^{(s)} - P^{(d)}, \label{eq:me_evolution}\\
\lbra \rho \rbra \frac{d \tilde{U}}{dt} &= MW + \langle \rho \rangle \epsilon^{(s)} + \langle \rho \rangle \epsilon^{(d)} - PD \label{eq:ie_evolution}.
\end{align}
Each of the sources in the evolution equations above is defined in \cref{tb:energy_sources}. We note that the derivation of the evolution equations for the four energies assumed a generic yet isotropic deformation tensor $G_{ij}$, and neglected the averaged heat flux since for homogeneous turbulence the averaged temperature is uniform in space \cite{blaisdell1991}. The intermode advection represents a transfer of energy from the solenoidal and dilatational modes, and thus satisfies $IA^{(s)} = -IA^{(d)}$. The production terms transfer the compression energy stored in the mean flow to the solenoidal and dilatational TKEs. The solenoidal and dilatational dissipations then transfer energy stored in the solenoidal and dilatational fields into heat. The pressure dilatation represents a two-way energy transfer between the mean internal energy and the dilatational TKE only, and the mechanical work transfers energy of the compression directly into heat. The mean-kinetic-energy advection and transport are not identically zero, unlike the case for the three other energies. All of these energy transfer mechanisms are depicted in \cref{fig:energy_chart}. We note that each energy component has a direct interaction with each of the other three energies. We also note that the driver for the interactions is the mean kinetic energy, since it has a predetermined time evolution that emulates the compression of the system. The other three energy components then respond in a self-consistent fashion to the time-evolution of the mean kinetic energy. The self-consistent feedback mechanism for the sudden viscous dissipation relies on these complex interactions, and thus can only be represented using the finite-Mach-number formulation and not the low-Mach-number assumption.

\section{Computational details}
\label{sec:comp_details}
Direct numerical simulations of \cref{eq:rho_miranda,eq:rhou_miranda,eq:rhoE_miranda} are carried out with the Miranda code developed at the Lawrence Livermore National Laboratory. This solver employs a tenth-order accurate Pad\'e scheme \cite{lele1992} for the discretization of the spatial derivatives, and a fourth-order, low-storage, five-step Runge-Kutta solver \cite{kennedy1999} for the temporal derivatives. An eighth-order compact filter is applied to the conserved variables $\rho$, $\rho u''_i$, and $E_t$ after each substep of the Runge-Kutta scheme, for the purposes of stability.

Miranda relies on the artificial-fluid-property approach to stabilize shock waves and contact discontinuities. Thus, an artificial bulk viscosity $\beta^*$ is introduced in the definition of the viscous stress tensor, and an artificial thermal conductivity $\kappa^*$ is added to the thermal conductivity $\kappa$ of the fluid. The artificial bulk viscosity and artificial thermal conductivity are computed as
\begin{align}
\beta^* &= \overline{ C_\beta \rho D(d)}, \\
\kappa^* &= \overline{ C_\kappa \rho \frac{c_v}{T \Delta t} D(T)}.
\end{align}
In the above, $\Delta t$ is the time step, the overbar denotes a truncated-Gaussian filter, and $D(\cdot)$ is an eighth-order derivative operator defined as
\begin{equation}
D(\cdot) = \max \left ( \left | \frac{\partial^8 \cdot}{\partial x^8} \right | \Delta x^{10}, \left | \frac{ \partial^8 \cdot}{\partial y^8} \right | \Delta y^{10}, \left | \frac{\partial^8 \cdot}{\partial z^8} \right | \Delta z^{10}  \right ).
\end{equation}
This operator strongly biases the artificial properties towards high wave numbers. The coefficients $C_\beta = 0.07$ and $C_\kappa = 0.001$ have been calibrated for simulations relevant to inertial confinement fusion, see for example \cite{weber2015,weber2014,weber2013}. For further details or capabilities of the code, the reader is referred to \cite{cook2004,cook2007,olson2014}.

The computational domain consists of a cube of length $2\pi$, with a uniform distribution of $256^3$ grid points. Periodic boundary conditions are applied on all sides of the cube. The ratio of specific heats has a value of $\gamma = 5/3$, and the Prandtl number is set to $Pr = 1$. The gas constant is computed as $R = R_u / M$, where the universal gas constant is $R_u = 8.314474 e10^7$ (cgs units), and the molar mass used is that of Deuterium, i.e.\@ $M = 2.014102$. Statistical quantities are obtained by averaging over all nodes of the mesh.

\label{sec:ie_results}
\begin{figure}[!t]
    \centering
    \includegraphics[width=0.49\textwidth]{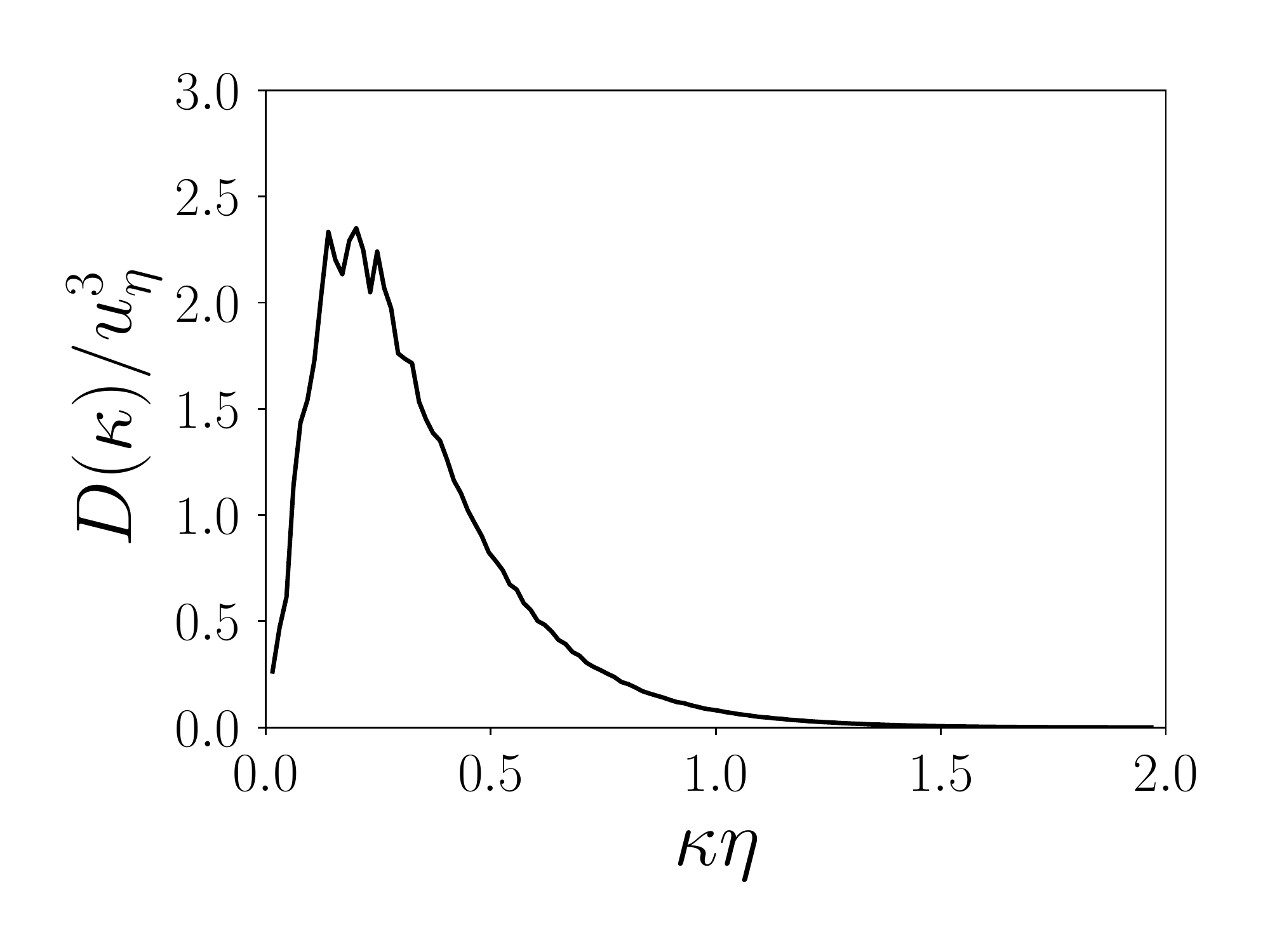}
    \caption{The dissipation spectrum, normalized by the Kolmogorov velocity $u_\eta = (\epsilon \nu)^{1/4}$, for simulations of linearly-forced compressible turbulence.}
    \label{fig:epsd_low_diss_spec}
\end{figure}

The initial flow field is extracted from a simulation of linearly-forced compressible turbulence \cite{petersen2010,campos2017}. This preliminary simulation is carried out for a duration of 18 initial eddy-turn-over times. The forcing coefficients introduced in \cite{petersen2010} require the specification of a priori values for the solenoidal and dilatational dissipations.  These two quantities were obtained from specifying a total dissipation $\epsilon = \epsilon^{(s)} + \epsilon^{(d)}$ and a dissipation ratio $\epsilon^{(d)} / \epsilon^{(s)}$. The value of the total dissipation is chosen a priori so that the corresponding Kolmogorov scale $\eta = (\nu^3 / \epsilon)^{1/4}$ is sufficiently large compared to the grid spacing. \Cref{fig:epsd_low_diss_spec} plots the dissipation spectrum at the end of the preliminary forced-turbulence simulation, and illustrates the range of scales resolved on the mesh, i.e.\@ $0 \le \kappa \eta \le 2$. This figure shows that we capture the long tail for high wavenumbers as it smoothly approaches a value of zero (a model spectrum for comparison is shown in \cite{pope2001}), and thus do not predict a fictitious energy pileup or unphysical rapid decay at the highest wave modes. This thus serves as further evidence that the combination of forcing coefficients and mesh resolution chosen appropriately capture all of the dissipative scales, as should be the case for any direct numerical simulation. The ratio of dissipations was set to $\epsilon^{(d)} / \epsilon^{(s)} = 0.01$. Simulations of compressing turbulence that were initialized from a linearly-forced case with $\epsilon^{(d)} / \epsilon^{(s)} = 1.0$ were also carried out. Results for this higher initial dissipation ratio are qualitatively similar to those of the lower dissipation ratio and are therefore not included in this paper. The turbulent Mach number and Taylor-scale Reynolds number are defined as
\begin{equation}
\label{eq:turb_mach_number}
M_t = \frac{\sqrt{ \widetilde{u''_i u''_i} }}{ \widetilde{c}},
\end{equation}
\begin{equation}
\label{eq:taylor_Re}
Re_\lambda = \left(\frac{20 k^2}{3 \epsilon \nu} \right)^{1/2},
\end{equation}
where $c = \sqrt{\gamma R T}$ is the speed of sound, and $\nu = \lbra \mu \rbra / \lbra \rho \rbra$ is the averaged kinematic viscosity. The extracted turbulent field at the end of the linear forcing has $M_t \approx 0.65$ and $Re_\lambda \approx 70$. The corresponding ratio of dilatational to solenoidal TKE is $k^{(d)} / k^{(s)} = 0.033$. For the linearly-forced simulations, the power law exponent is set to the traditional fluid value of $n = 3/4$. However, once the isotropic compression is applied to the initial flow field, the power law exponent is switched to the value used in \cite{davidovits2016}, namely $n= 5/2$, so as to reproduce the sudden viscous dissipation mechanism. The initial ratio of the artificial dissipation \cite{campos2017} to physical dissipation is 0.016, which decays rapidly as the compression is initiated. Thus, the simulations are mostly affected by physical rather than artificial dissipative mechanisms, as should be the case for a properly-refined direct numerical simulation \cite{olson2014}.


\section{Results}
\label{sec:results}
The analysis of the self-consistent feedback mechanism is divided into two subsections. The first focuses on the behavior of the TKEs and the various mechanisms depicted in \cref{fig:energy_chart} that modulate their temporal evolution. The second half of the analysis is centered around the resulting evolution of the internal energy, and the amplification of the temperature due to the viscous dissipation.

\subsection{Turbulent kinetic energies}
\label{sec:tke_results}
\subsubsection{Profile histories}

\begin{figure}[!t]
    \centering
    \begin{subfigure}[b]{0.49\textwidth}
        \includegraphics[width=\textwidth]{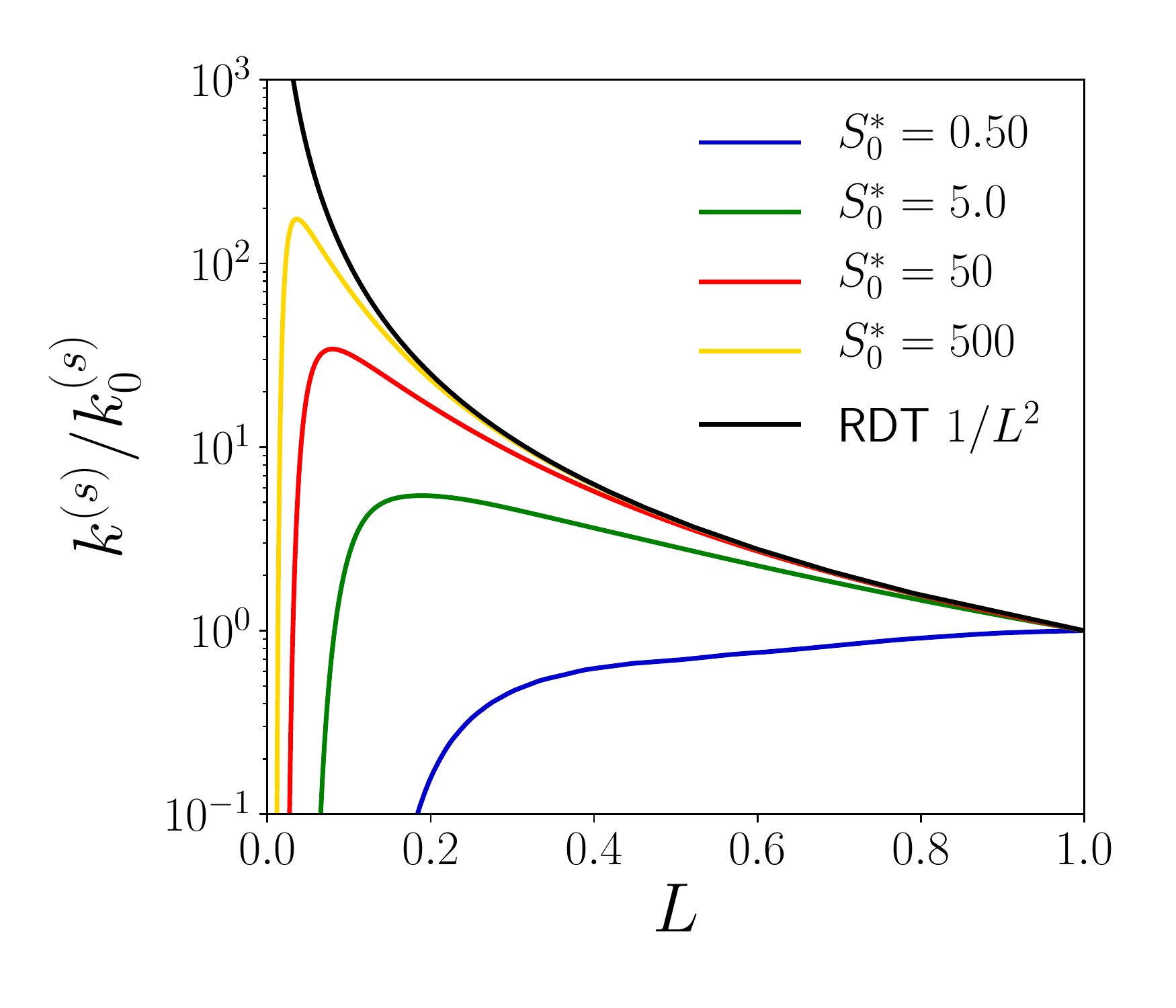}
        \caption{}
        \label{fig:epsd_low_tkes}
    \end{subfigure}
    \begin{subfigure}[b]{0.49\textwidth}
        \includegraphics[width=\textwidth]{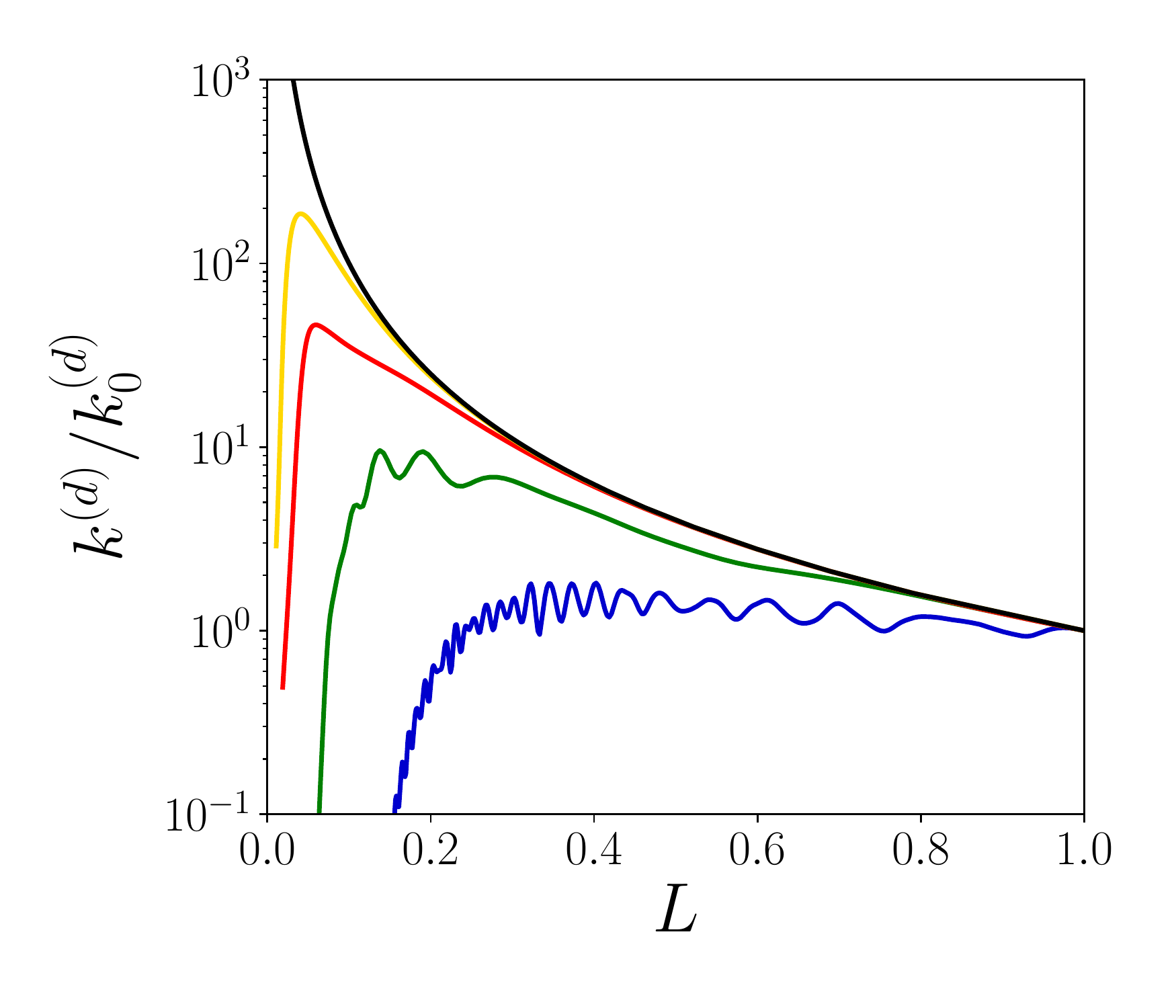}
        \caption{}
        \label{fig:epsd_low_tked}
    \end{subfigure}
    \caption{Evolution of (a) solenoidal and (b) dilatational TKE as a function of the size of the domain $L$. The solenoidal and dilatational TKEs are normalized by their initial values $k^{(s)}_0$ and $k^{(d)}_0$, respectively. The initial length of the domain is $L=1$, which decreases as time progresses. The colors correspond to different values of $S^*_0$, which is the shear parameter $S^*$ at time $t=0$.}
    \label{fig:epsd_low_tke}
\end{figure}

The time evolution of the solenoidal and dilatational TKEs are shown in \cref{fig:epsd_low_tkes,fig:epsd_low_tked}, respectively. As done in \cite{davidovits2016}, rather than plotting the TKE evolution against time, a parameterization in terms of the length of the domain $L$ is used, and thus time progresses from right to left. Also equivalent to the results in \cite{davidovits2016}, the TKE evolutions for different compression speeds $\dot{L}$ are shown. These different cases are labelled by the initial value of the RDT parameter $ S^* = S k / \epsilon $, where $S = \dot{L} / L$ is the inverse time scale of the compression, and $ k / \epsilon$ is the time scale of the turbulence. For sufficiently large values of $S^*$, the compression is rapid enough that the non-linear turbulence-turbulence interactions are negligible, and the evolution of the turbulence is described exactly by rapid-distortion theory (RDT) \cite{pope2001,blaisdell1996}.

As \cref{fig:epsd_low_tkes} shows, the evolution of solenoidal TKE exhibits the sudden viscous dissipation mechanism of \cite{davidovits2016}. Even though the compression speeds used in this study are different from those of \cite{davidovits2016}, there is strong qualitative agreement between the results shown in \cref{fig:epsd_low_tkes} and those in fig.\@ 1 of \cite{davidovits2016}. \Cref{fig:epsd_low_tked} shows that the dilatational TKE also exhibits the sudden viscous dissipation mechanism. Additionally, strong agreement with RDT \cite{blaisdell1996} is observed for both the solenoidal and dilatational modes given the fastest compression rates. A few differences between the behavior of solenoidal and dilatational TKE are noted. First, increasingly strong oscillations of dilatational TKE are observed as $S^*_0$ is decreased. This highly oscillatory behavior is discussed further in \cref{sec:tke_budget}. Second, the dilatational energy lags the solenoidal energy. At the last recorded instance in time, the solenoidal TKE has decayed by more than two orders of magnitude for case $S^*_0 = 50$, whereas the dilatational TKE has decayed by about two orders of magnitude only. For the $S^*_0 = 500$ case, the solenoidal TKE has decayed by more than three orders of magnitude and the dilatational TKE has decreased by less than two orders of magnitude. Lastly, for the slowest compression rate, the solenoidal and dilatational TKE diverge in their initial behavior: whereas the dilatational TKE slightly increases until it suddenly dissipates, the solenoidal TKE decreases from the start. 

\begin{figure}[!t]
    \centering
    \begin{subfigure}[b]{0.49\textwidth}
        \includegraphics[width=\textwidth]{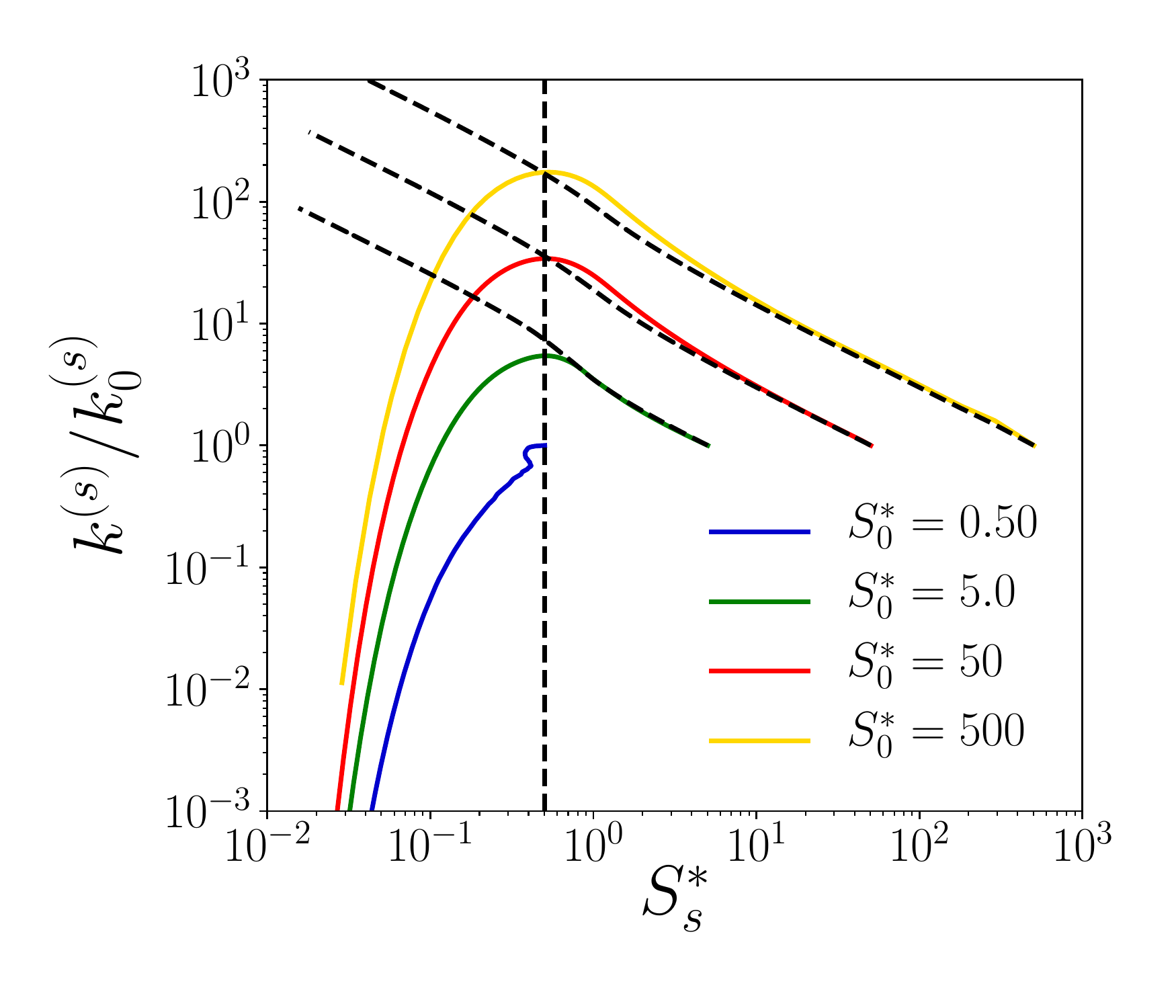}
        \caption{}
        \label{fig:epsd_low_Ske_s}
    \end{subfigure}    
    \begin{subfigure}[b]{0.49\textwidth}
        \includegraphics[width=\textwidth]{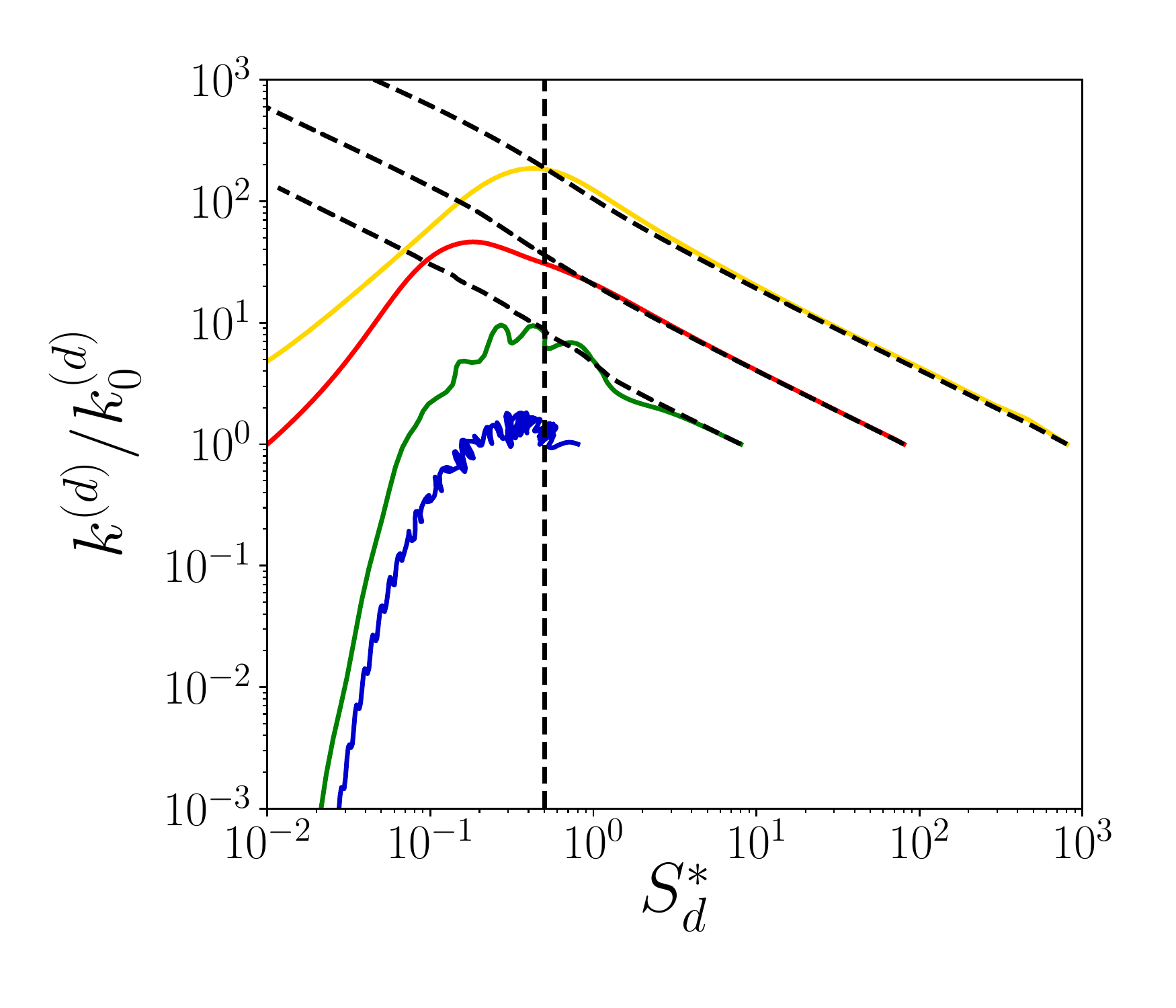}
        \caption{}
        \label{fig:epsd_low_Ske_d}
    \end{subfigure}
    \caption{Profiles of (a) solenoidal and (b) dilatational TKE plotted against the shear parameter $S^*_\alpha = S k^{(\alpha)} / \epsilon^{(\alpha)}$, where $\alpha = s$ for fig.\@ (a) and $\alpha = d$ for fig.\@ (b). The solenoidal and dilatational TKEs are normalized by their initial values $k^{(s)}_0$ and $k^{(d)}_0$, respectively. The vertical dashed line in the left plot corresponds to the time at which $P^{(s)} = \lbra \rho \rbra \epsilon^{(s)}$, and the vertical dashed line on the right plot corresponds to the time at which $P^{(d)} = \lbra \rho \rbra \epsilon^{(d)}$. The dashed diagonal lines correspond to \cref{eq:Ske_s_scaling}.}
    \label{fig:epsd_low_Ske}
\end{figure}

An alternate representation of the evolution of TKEs is shown in \cref{fig:epsd_low_Ske_s,fig:epsd_low_Ske_d}. In \cref{fig:epsd_low_Ske_s} the evolution of the solenoidal TKE is parameterized by the solenoidal shear parameter $S^*_s = S k^{(s)}/\epsilon^{(s)}$, and in \cref{fig:epsd_low_Ske_d} the evolution of the dilatational TKE is parameterized by the dilatational shear parameter $S^*_d = S k^{(d)}/\epsilon^{(d)}$. The dashed vertical lines correspond to the point in time at which production is equal to dissipation, that is, the point at which $P^{(s)} = \lbra \rho \rbra \epsilon^{(s)}$ for \cref{fig:epsd_low_Ske_s} and $P^{(d)} = \lbra \rho \rbra \epsilon^{(d)}$ for \cref{fig:epsd_low_Ske_d}. The dashed diagonal lines correspond the assumption of RDT scaling, for which $\epsilon^{(\alpha)} \sim \left (k^{(\alpha)} \right)^3$ for $\alpha = s,d$. That is, assuming this relationship between TKE and dissipation, the TKE can be expressed as
\begin{equation}
\label{eq:Ske_s_scaling}
k^{(\alpha)} = \frac{\left ( k^{(\alpha)} \right)^{3/2} }{ \left ( k^{(\alpha)} \right)^{1/2} } \sim \frac{ \left ( \epsilon^{(\alpha)} \right)^{1/2} }{ \left ( k^{(\alpha)} \right)^{1/2} } \sim S^{1/2} \left . S_\alpha^* \right .^{-1/2}
\end{equation}
for $\alpha = s,d$. It is important to note that $k^{(\alpha)}$ does not scale simply as $\left. S_\alpha^* \right .^{-1/2}$ since $S$ also depends on time. However, the dependence of $S$ on time is given by the predetermined and known compression history of the domain $L$.

\Cref{fig:epsd_low_Ske_s} shows that for compressions $S^*_0 = 5.0$, 50, and $500$, the increase in solenoidal TKE is in close agreement with RDT. That is, the RDT assumption of $\epsilon^{(\alpha)} \sim \left (k^{(\alpha)} \right)^3$ holds quite well during the initial phase of the compression. The agreement with the scaling of \cref{eq:Ske_s_scaling} could be beneficial for modeling purposes. Significant divergence from the RDT scaling occurs once the vertical line at which solenoidal production equals solenoidal dissipation is reached. After this point, the solenoidal dissipation overtakes the solenoidal production, and the turbulence decays. For the $S^*_0 = 0.50$ case, the compression is slow enough that the solenoidal production is never larger than the solenoidal dissipation, and thus the entire curve is located to the left of the vertical dashed line. \Cref{fig:epsd_low_Ske_d} shows a similar trend. We first note that the rapid oscillations in the dilatational profiles corresponding to slow compression speeds are also evident in this figure. The agreement with the scaling $\epsilon^{(d)} \sim \left ( k^{(d)} \right)^3$ still holds for the $S^*_0 = 5.0, 50,\text{ and }500$ cases, although, for the $S^*_0 = 5.0$ case, this agreement is not as strong as that of the corresponding solenoidal field. More importantly, the vertical line at which dilatational production equals dilatational dissipation no longer demarcates the domains of increasing and decreasing turbulence for all four cases, since for $S^*_0 = 50$ the dilatational TKE keeps on increasing after this vertical line is reached. Lastly, \cref{fig:epsd_low_Ske_d} shows that the decrease in energy is slower than that observed in \cref{fig:epsd_low_Ske_s} for cases $S^*_0 = 50$ and 500. This suggests that the dilatational dissipation is acting against an additional source, which, as will be shown in \cref{sec:tke_budget}, turns out to be the pressure dilatation.

\subsubsection{Budgets}
\label{sec:tke_budget}

\begin{figure}
    \centering
    \begin{subfigure}[b]{0.49\textwidth}
        \includegraphics[width=\textwidth]{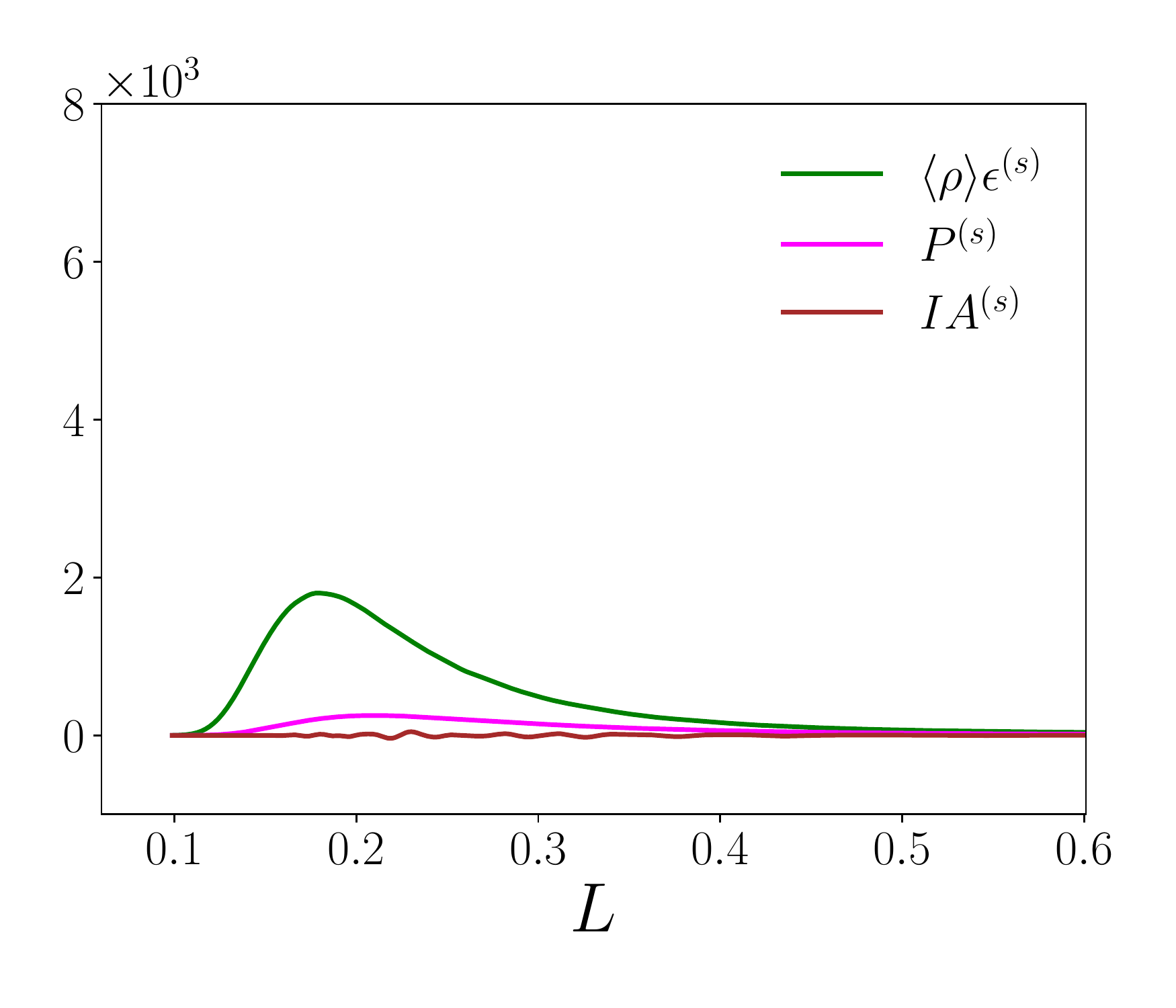}
        \label{fig:epsd_low_ks_bal_1}
    \end{subfigure}
    \begin{subfigure}[b]{0.49\textwidth}
        \includegraphics[width=\textwidth]{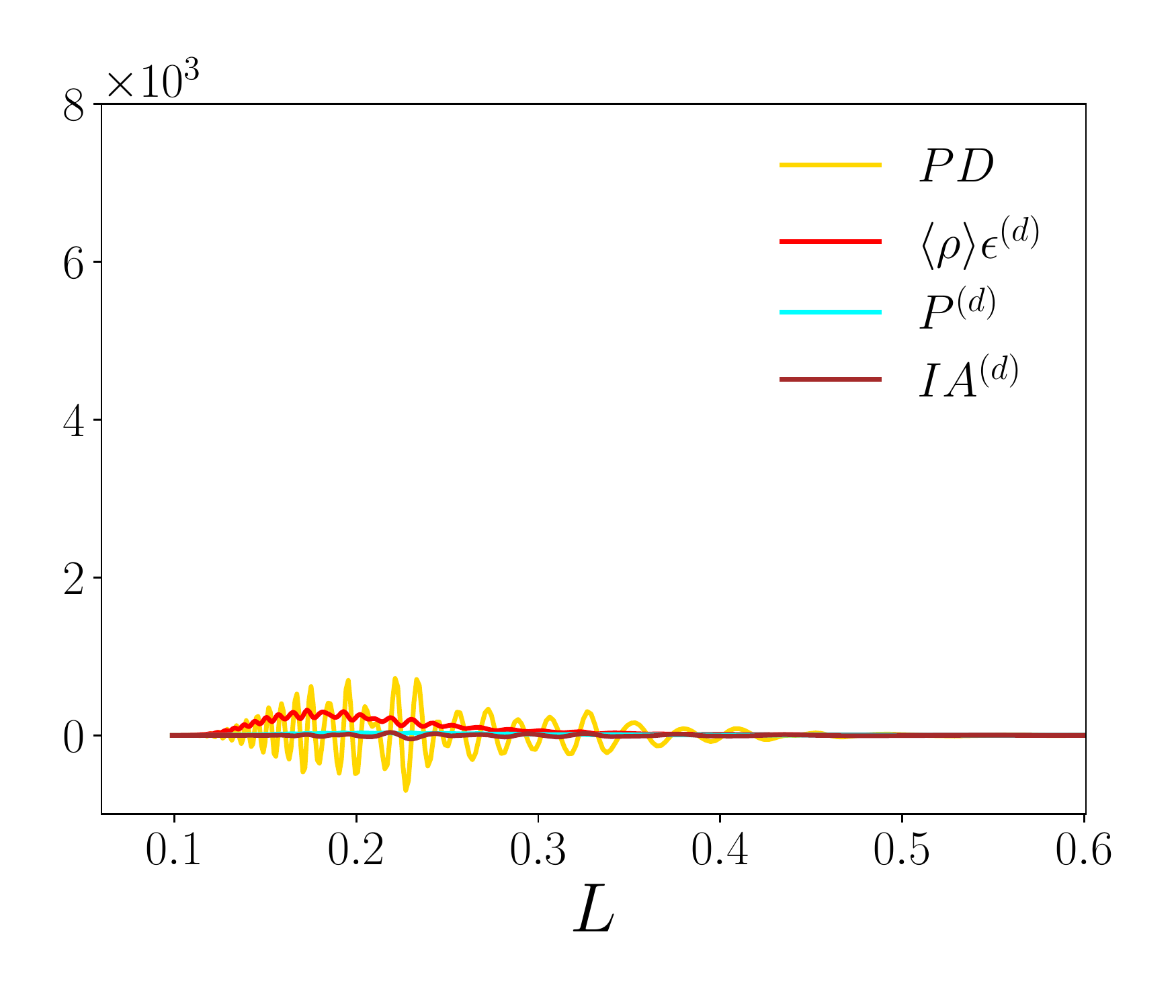}
        \label{fig:epsd_low_kd_bal_1}
    \end{subfigure} 
    \caption{TKE budget for the solenoidal mode on the left and the dilatational mode on the right, for $S^*_0 = 0.5$. The initial length of the domain is $L=1$, which decreases as time progresses. All terms have been normalized by $\rho_0 U_0^3 / L_0$.}
    \label{fig:epsd_low_tke_budget_1}
\end{figure}

\begin{figure}
    \centering
    \begin{subfigure}[b]{0.49\textwidth}
        \includegraphics[width=\textwidth]{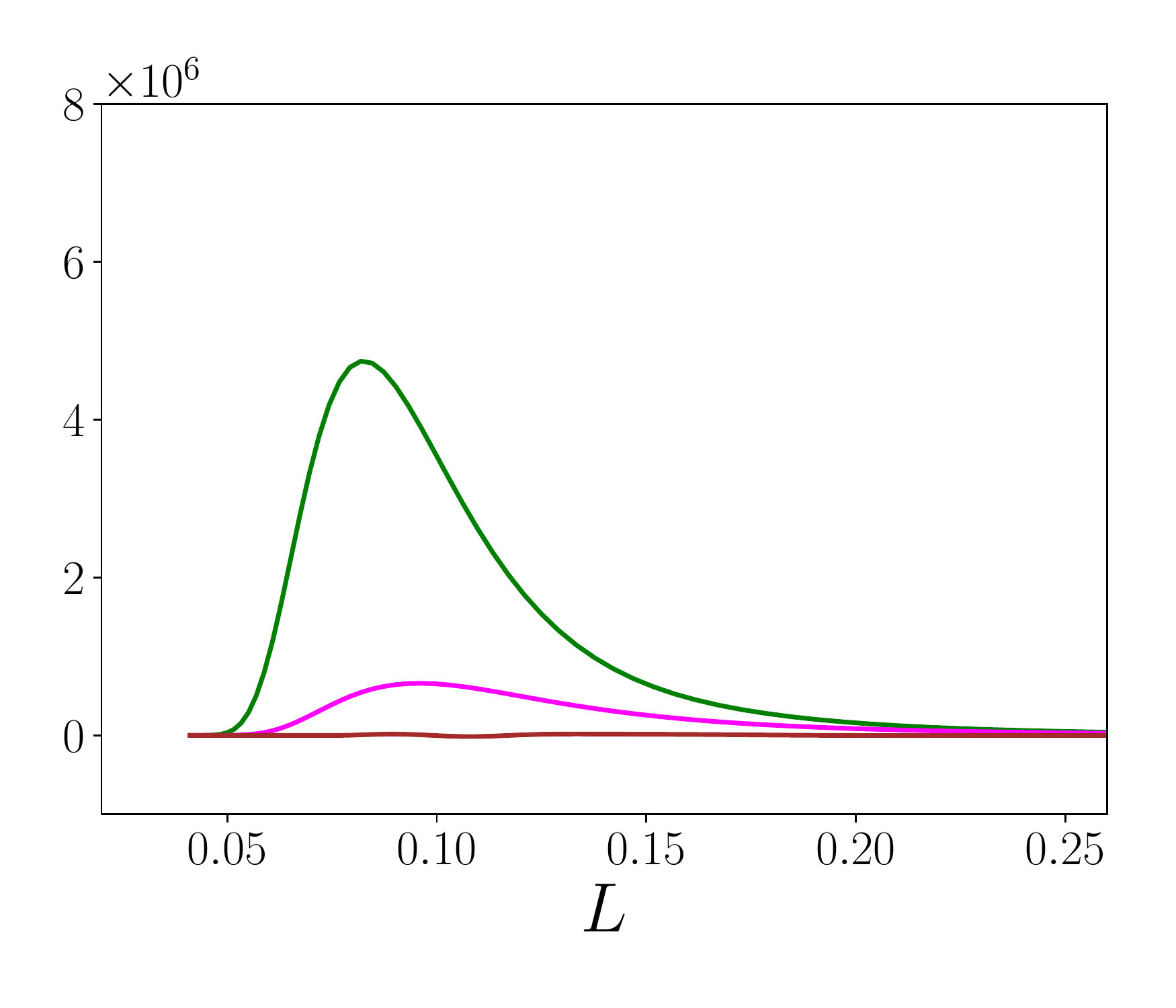}
        \label{fig:epsd_low_ks_bal_2}
    \end{subfigure}
    \begin{subfigure}[b]{0.49\textwidth}
        \includegraphics[width=\textwidth]{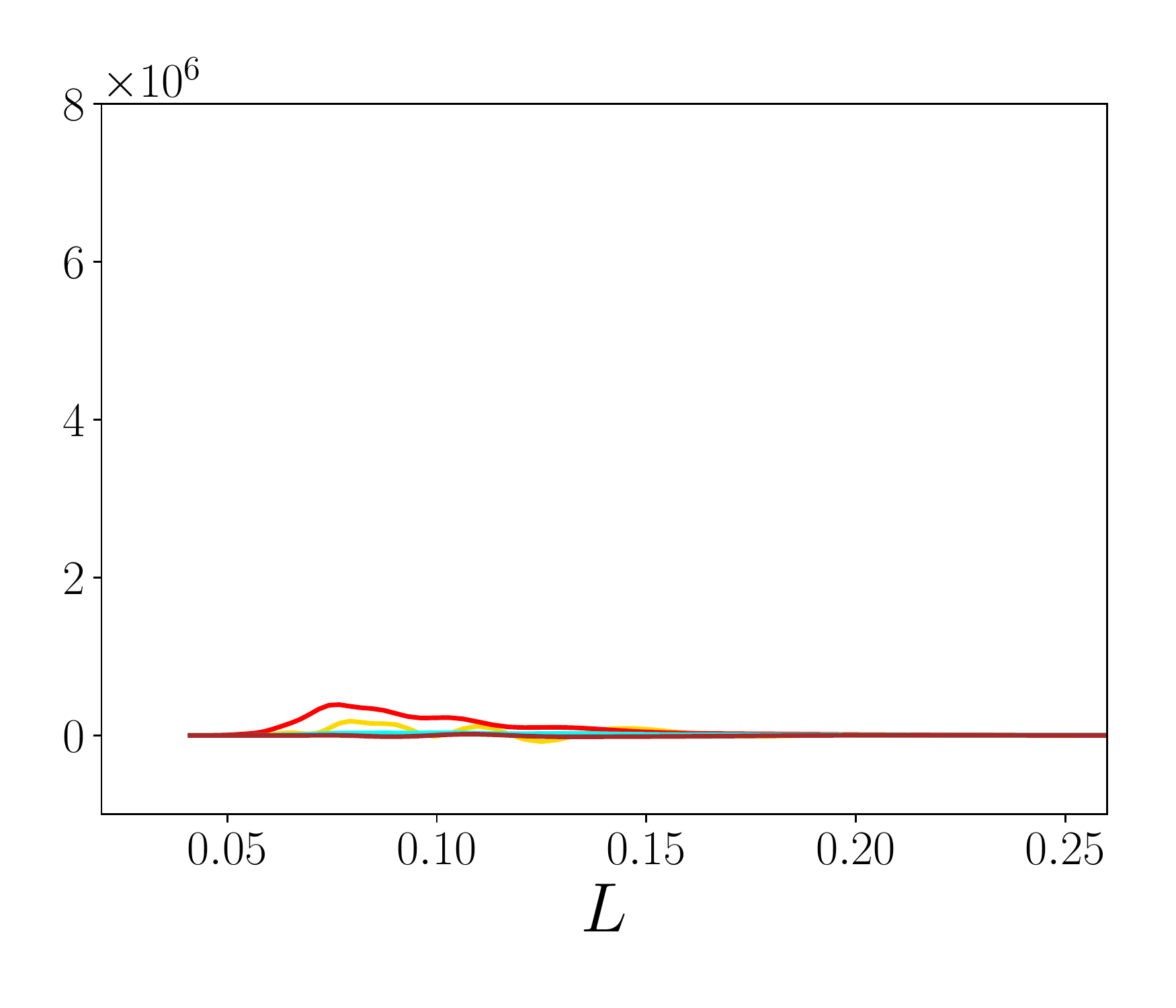}
        \label{fig:epsd_low_kd_bal_2}
    \end{subfigure} 
    \caption{TKE budget for the solenoidal mode on the left and the dilatational mode on the right, for $S^*_0 = 5.0$. The initial length of the domain is $L=1$, which decreases as time progresses. All terms have been normalized by $\rho_0 U_0^3 / L_0$. The same legend as that of \cref{fig:epsd_low_tke_budget_1} applies to the plots above.}
    \label{fig:epsd_low_tke_budget_2}
\end{figure}

\begin{figure}
    \centering
    \begin{subfigure}[b]{0.49\textwidth}
        \includegraphics[width=\textwidth]{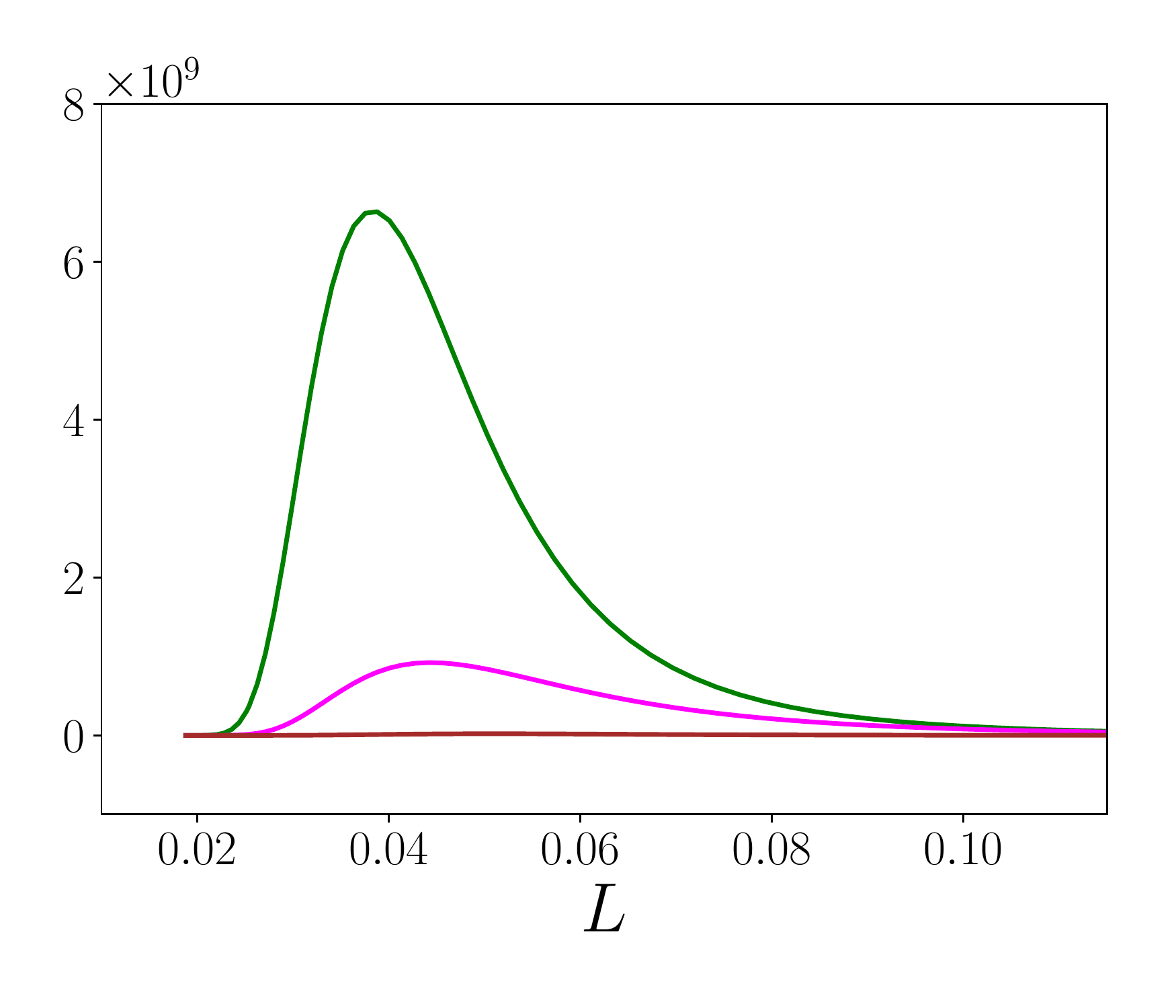}
        \label{fig:epsd_low_ks_bal_3}
    \end{subfigure}
    \begin{subfigure}[b]{0.49\textwidth}
        \includegraphics[width=\textwidth]{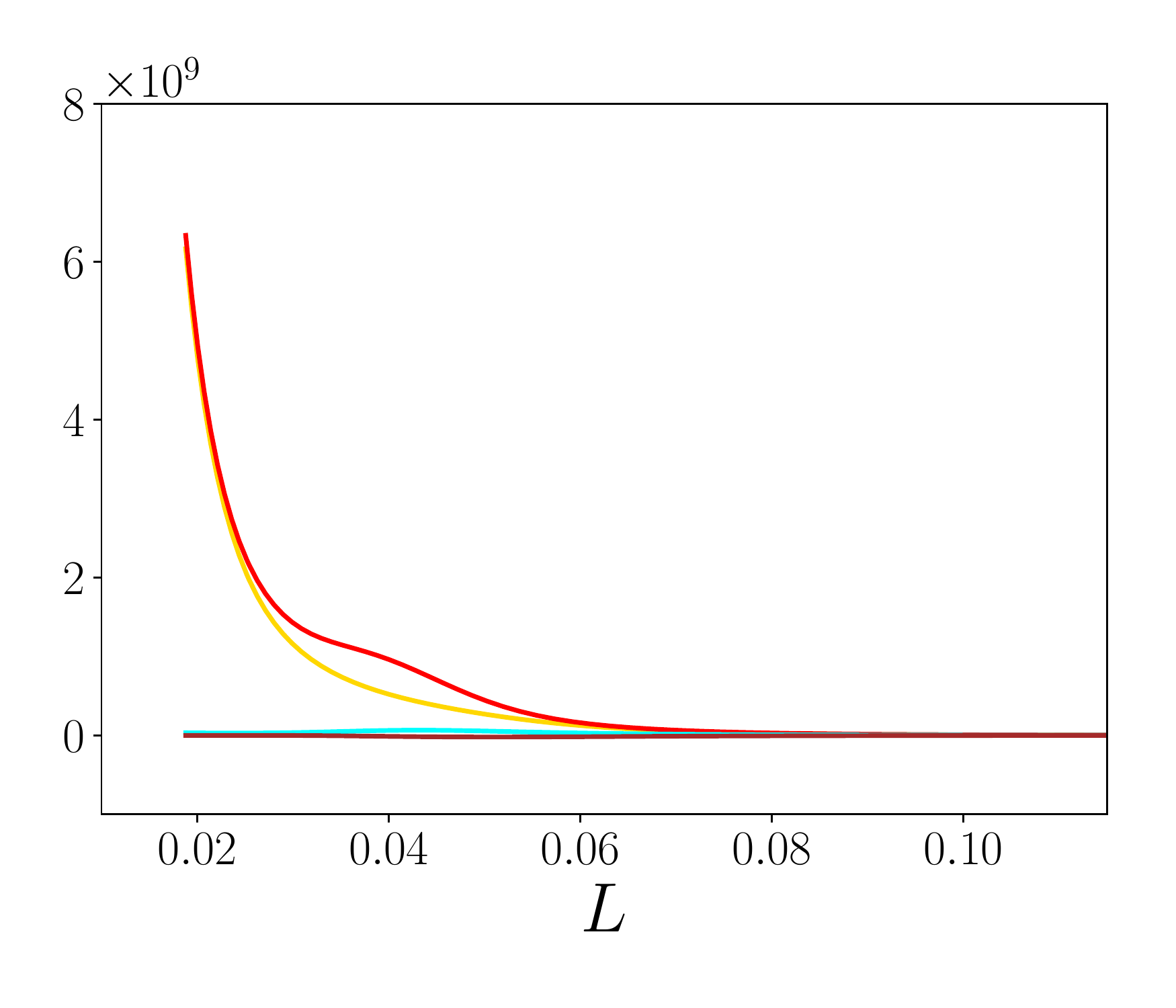}
        \label{fig:epsd_low_kd_bal_3}
    \end{subfigure}
    \caption{TKE budget for the solenoidal mode on the left and the dilatational mode on the right, for $S^*_0 = 50$. The initial length of the domain is $L=1$, which decreases as time progresses. All terms have been normalized by $\rho_0 U_0^3 / L_0$. The same legend as that of \cref{fig:epsd_low_tke_budget_1} applies to the plots above.}
    \label{fig:epsd_low_tke_budget_3}
\end{figure}

\begin{figure}
    \centering
    \begin{subfigure}[b]{0.49\textwidth}
        \includegraphics[width=\textwidth]{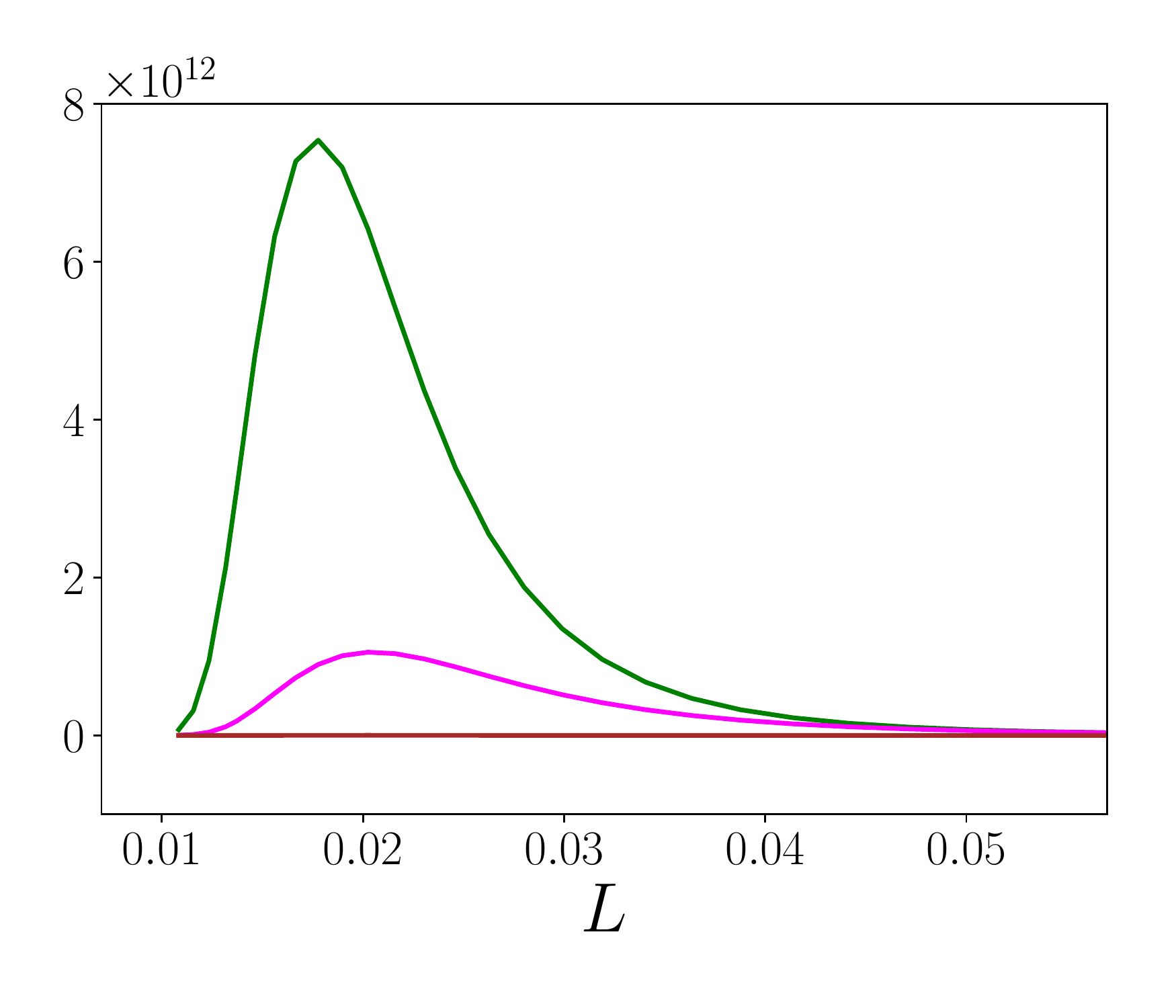}
        \label{fig:epsd_low_ks_bal_4}
    \end{subfigure}
    \begin{subfigure}[b]{0.49\textwidth}
        \includegraphics[width=\textwidth]{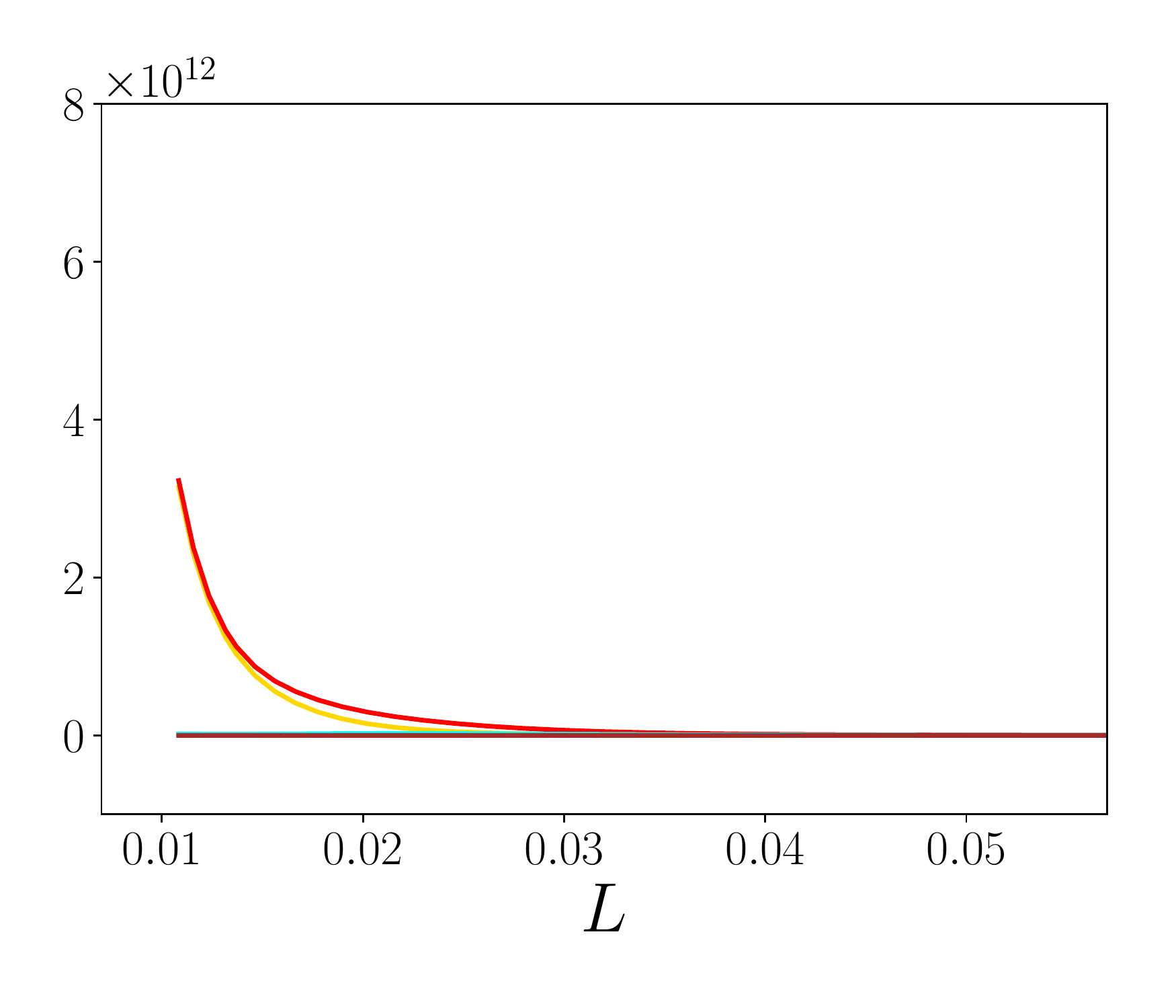}
        \label{fig:epsd_low_kd_bal_4}
    \end{subfigure}
    \caption{TKE budget for the solenoidal mode on the left and the dilatational mode on the right, for $S^*_0 = 500$. The initial length of the domain is $L=1$, which decreases as time progresses. All terms have been normalized by $\rho_0 U_0^3 / L_0$. The same legend as that of \cref{fig:epsd_low_tke_budget_1} applies to the plots above.}
    \label{fig:epsd_low_tke_budget_4}
\end{figure}

\Cref{fig:epsd_low_tke_budget_1,fig:epsd_low_tke_budget_2,fig:epsd_low_tke_budget_3,fig:epsd_low_tke_budget_4} contain the TKE budget for the solenoidal and dilatational fields. For the $S^*_0 = 0.5$ case shown in \cref{fig:epsd_low_tke_budget_1}, oscillations in the pressure dilatation and dilatational dissipation are observed. The magnitude of the oscillations in $\lbra \rho \rbra \epsilon^{(d)}$ are significantly smaller than those of PD. We also note that the oscillations of the pressure dilatation and dilatational dissipation are correlated, with the dilatational dissipation slightly lagging the pressure dilatation. Moreover, the oscillations in $k^{(d)}$ shown in \cref{fig:epsd_low_tked} are also correlated with  $PD$, with $k^{(d)}$ lagging behind PD. This serves as evidence that pressure dilatation is responsible for the oscillatory behavior of the dilatational TKE. The strong oscillatory nature of $PD$ has been observed elsewhere, see for example \cite{kida1990,miura1995} for the case of forced turbulence and \cite{blaisdell1991,sarkar1992} for sheared turbulence. For the $S^*_0 = 5.0$ case shown in \cref{fig:epsd_low_tke_budget_2} the oscillations in $PD$ and $\lbra \rho \rbra \epsilon^{(d)}$ have been attenuated. \Cref{fig:epsd_low_tke_budget_3,fig:epsd_low_tke_budget_4} show that as the compression speed is increased to $S^*_0 = 50$ and 500, $PD$ and $\lbra \rho \rbra \epsilon^{(d)}$ do not exhibit oscillations up to the last simulated instance in time. 

Some trends shown by \cref{fig:epsd_low_tke_budget_1,fig:epsd_low_tke_budget_2,fig:epsd_low_tke_budget_3,fig:epsd_low_tke_budget_4} are important to be noted. As the compression speed increases, the peak values of the profiles occur at smaller and smaller domain lengths, as highlighted by the different ranges used for the $x$ axis. This is in agreement with \cref{fig:epsd_low_tke} and the results in \cite{davidovits2016}. However, more relevant to the current study is that, as the compression speed increases, peak values for the dilatational TKE sources occur at even smaller domain lengths than those of the solenoidal energy. For example, for compression speeds $S^*_0 = 50$ and 500, peak values for the solenoidal dissipation and production are reached before those of dilatational dissipation and pressure dilatation. These last two terms begin to increase rapidly only after the solenoidal dissipation and production are already decaying from their maximum values. This lag in the dilatational modes with respect to the solenoidal ones was also observed in \cref{fig:epsd_low_tke} as previously noted, but is now more apparent due to the smaller scale of the $x$ axis in \cref{fig:epsd_low_tke_budget_1,fig:epsd_low_tke_budget_2,fig:epsd_low_tke_budget_3,fig:epsd_low_tke_budget_4}. The second trend to highlight is that, as the compression speed is increased, the solenoidal sources do not show as distinct a change in behavior as the dilatational sources. The solenoidal dissipation and production profiles retain the same shape, though with different magnitudes, for the different compression rates. On the other hand, the pressure dilatation and dilatational dissipation go from a highly-oscillatory pattern to smoothly-varying non-oscillatory profiles for the fastest compression speed. This poses formidable challenges for modeling purposes. The last trend to highlight is that the pressure dilatation is either skewed towards positive values, as is the case for $S^*_0 = 0.5$, or is positive throughout the entire compression. This is further exemplified by looking at \cref{tb:energy_integrated_sources}, which shows the integrated values of the energy transfer mechanisms, from the initial to the last available simulated time. All integrated values for the pressure dilatation are positive. Thus, $PD$ behaves more as a source rather than a sink or a neutral term in the balance of dilatational TKE. Thus, the dilatational dissipation needs to counteract the effect of both the dilatational production and pressure dilatation for the sudden viscous dissipation to occur in the dilatational field. Given that for the two fastest compressions the integrated contribution of $PD$ is almost as large as that of $\lbra \rho \rbra \epsilon^{(d)}$, it is thus not unexpected that the dilatational TKE decays at a slower rate than the solenoidal TKE, as shown in \cref{fig:epsd_low_Ske}. 


\begin{table}
\centering
\caption{Integrated energy sources given different compression speeds. All values are normalized by $\rho_0 U_0^2$.}
\label{tb:energy_integrated_sources}
\begin{tabular}{ Sc Sc Sc Sc Sc}
 & $S^*_0 = 0.5$ & $S^*_0 = 5.0$ & $S^*_0 = 50$ & $S^*_0 = 500$ \\
\hline\hline
$IA^{(s)}$ & 2.32e-01 & 4.80e+01& 4.45e+03 & 4.90e+03 \\
$IA^{(d)}$ & -2.32e-01& -4.80e+01 & -4.45e+03 & -4.90e+03 \\
$P^{(s)}$ & 2.89e+01& 3.08e+03 & 1.97e+05 & 1.03e+07 \\
$P^{(d)}$ & 2.80e+00 & 1.57e+02 & 1.39e+04 & 3.23e+05 \\
$\lbra \rho \rbra \epsilon^{(s)}$ & 1.46e+02& 1.54e+04 & 9.82e+05 & 5.07e+07 \\
$\lbra \rho \rbra \epsilon^{(d)}$ & 1.87e+01& 1.07e+03 & 3.24e+05 & 6.59e+06 \\
$PD$ & 4.89e+00& 3.34e+02 & 2.64e+05 & 5.15e+06 \\
$MW$ & 3.34e+05& 2.69e+07 & 1.35e+09 & 2.14e+10 \\
\end{tabular}
\end{table}

\subsubsection{Spectra}
\begin{figure}[!t]
    \centering
    \begin{subfigure}[b]{0.49\textwidth}
        \includegraphics[width=\textwidth]{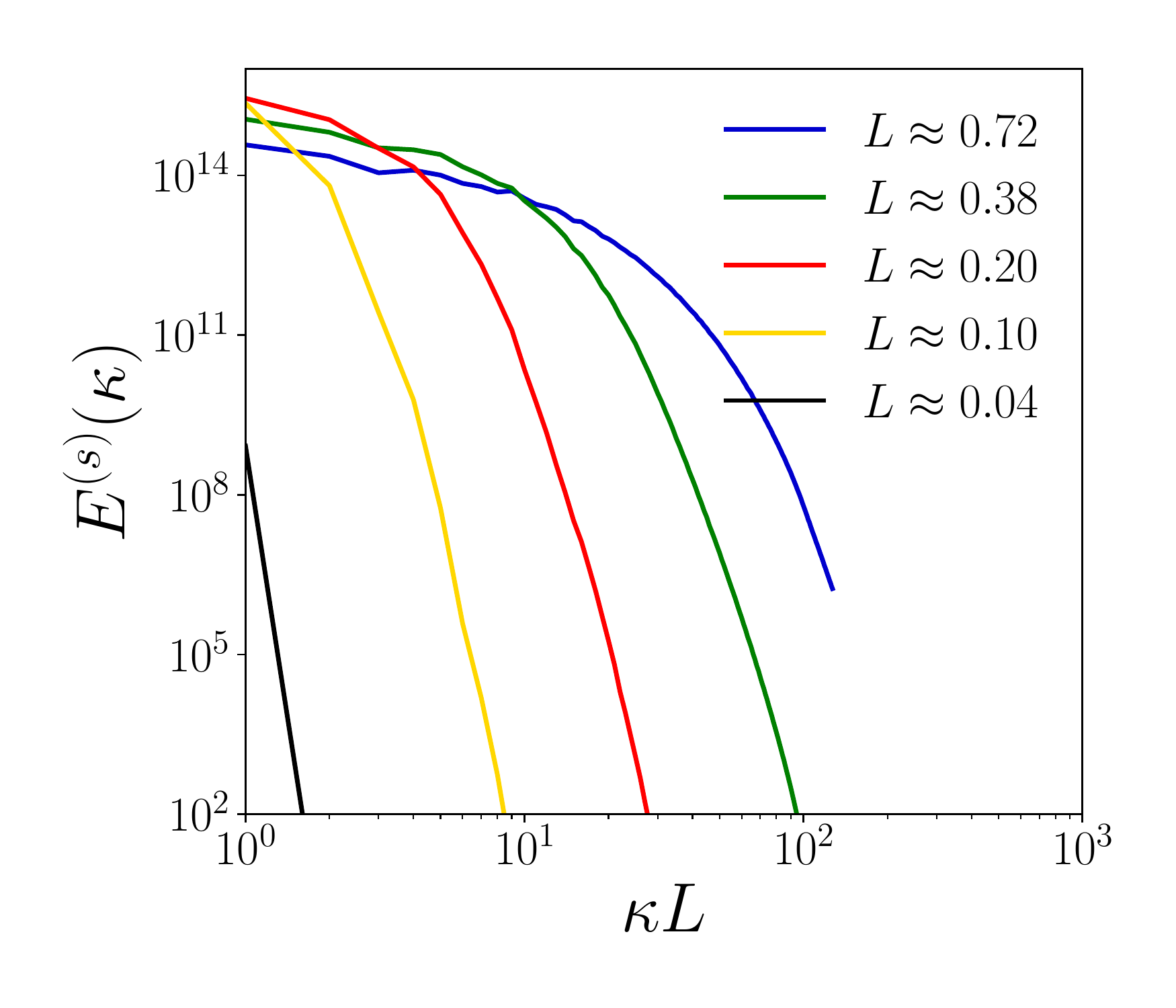}
        \caption{}
        \label{fig:epsd_low_spec_tkes}
    \end{subfigure}
    \begin{subfigure}[b]{0.49\textwidth}
        \includegraphics[width=\textwidth]{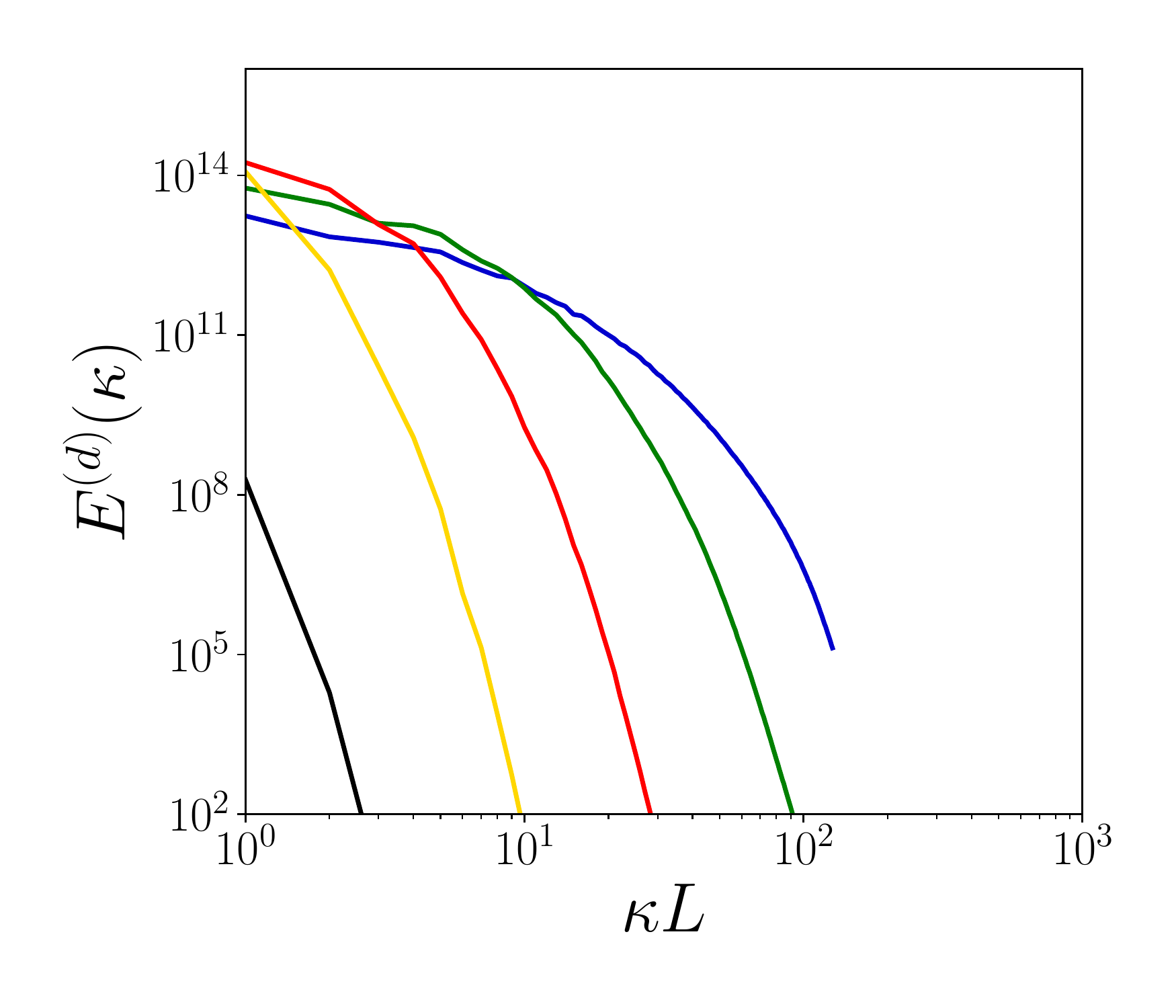}
        \caption{}
        \label{fig:epsd_low_spec_tked}
    \end{subfigure}
    \caption{Energy spectra for the (a) solenoidal and (b) dilatational TKE, at different times (or domain lengths) throughout the compression. The spectra correspond to the $S^*_0 = 5.0$ case.}
    \label{fig:epsd_low_tke_spectra}
\end{figure}

The energy spectra for the solenoidal and dilatational fields are shown in \cref{fig:epsd_low_tke_spectra}, for the compression speed of $S^*_0 = 5.0$. The profile obtained at $L \approx 0.10$ corresponds to a point in time for which the sudden viscous dissipation mechanism is taking place, and the profile at $L \approx 0.04$ to a time for which most of the turbulence has already been dissipated. The shapes and trends exhibited by the solenoidal and dilatational spectra are equal to each other. Additionally, these profiles are in qualitative agreement with results shown in \cite{davidovits2016}. As the compression proceeds, the energy in the higher modes decreases whereas the energy in the lower modes increases. The set of modes for which the energy decreases expands as the compression progresses, and eventually even the lower modes are dissipated, as shown by the profile at $L \approx 0.04$. 

\subsection{Internal energy}
\label{sec:ie_results}
\begin{figure}[!t]
    \centering
    \includegraphics[width=0.49\textwidth]{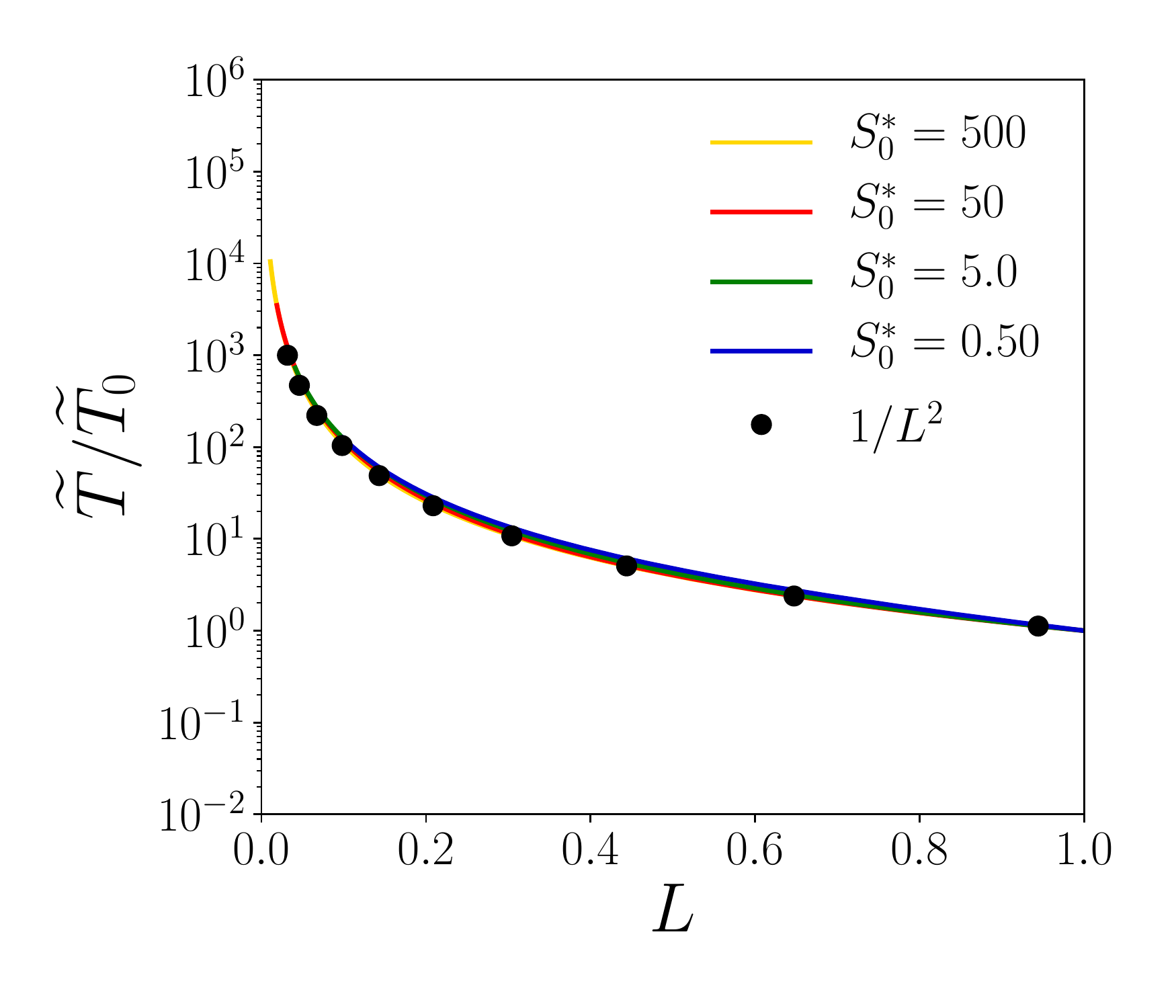}
    \caption{Evolution of temperature as a function of the size of the domain. The initial temperature is denoted by $\widetilde{T}_0$.}
    \label{fig:epsd_low_T}
\end{figure}

\begin{figure}
    \centering
    \begin{subfigure}[b]{0.49\textwidth}
        \includegraphics[width=\textwidth]{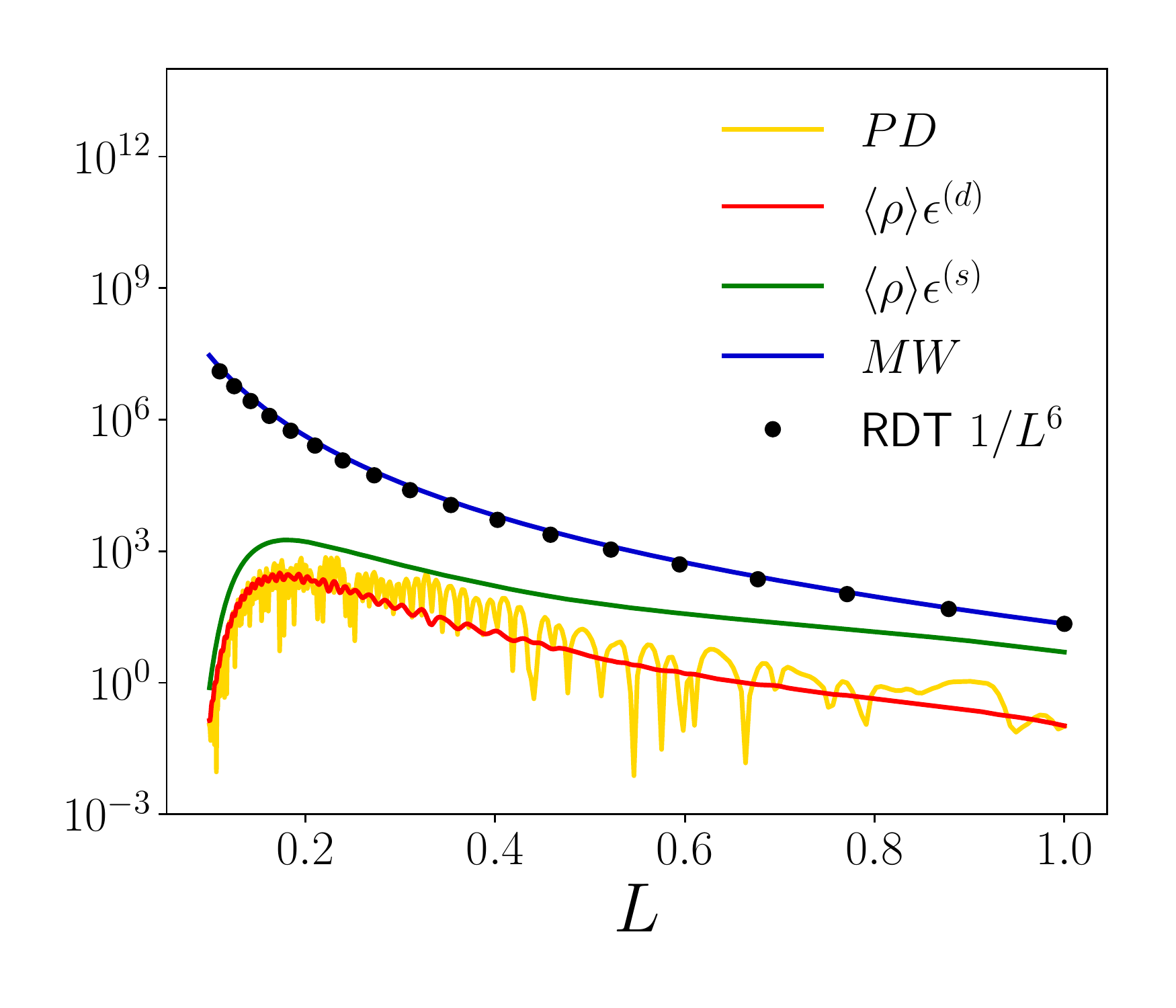}
        \label{fig:epsd_low_ie_bal_log_1}
    \end{subfigure}
    \begin{subfigure}[b]{0.49\textwidth}
        \includegraphics[width=\textwidth]{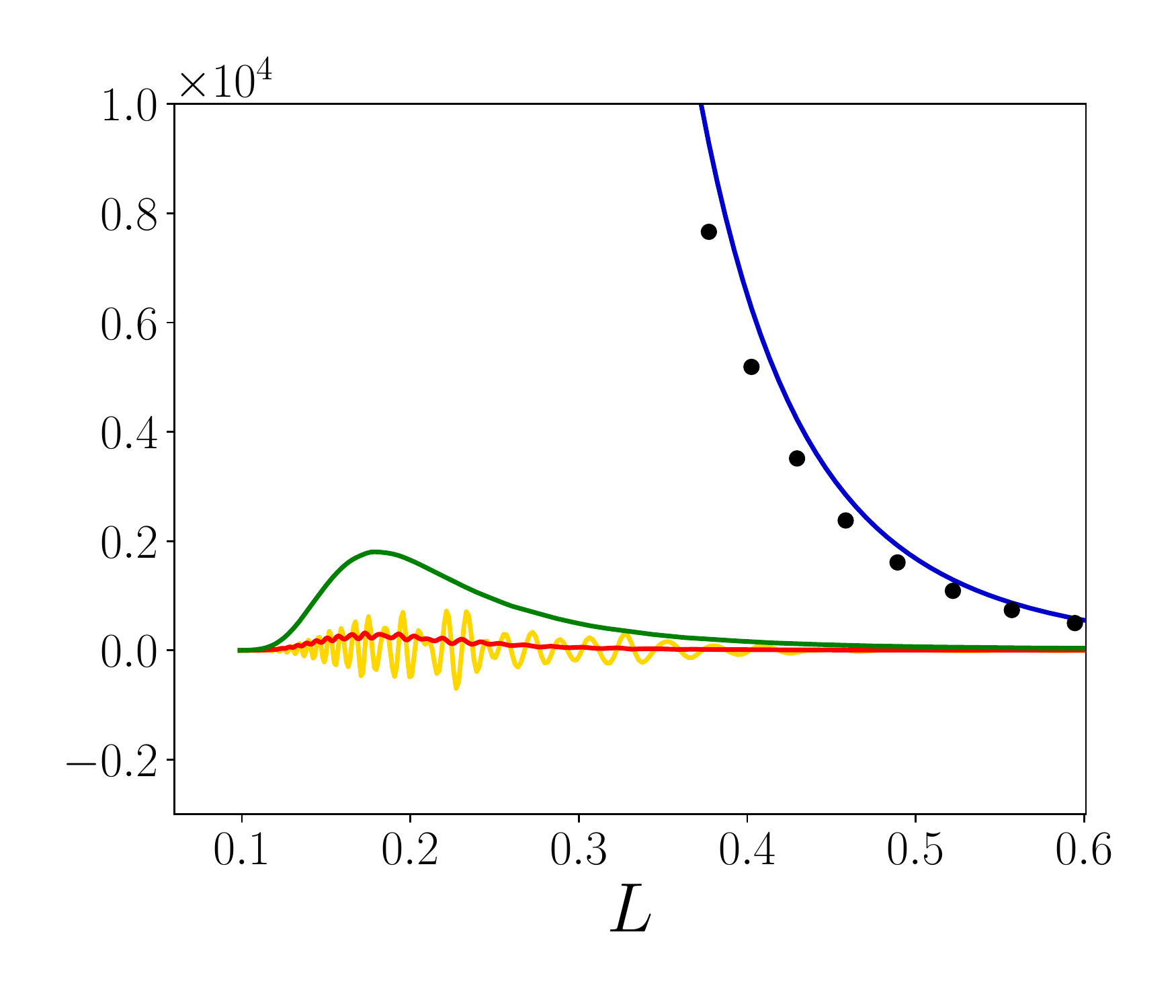}
        \label{fig:epsd_low_ie_bal_linear_1}
    \end{subfigure}   
    \caption{Mean internal energy budget in log scale on the left and linear scale on the right, for $S^* = 0.50$. The initial length of the domain is $L=1$, which decreases as time progresses. All terms have been normalized by $\rho_0 U_0^3 / L_0$. The absolute value of $PD$ is shown for the log-scale plot.}
    \label{fig:epsd_low_ie_budget_1}
\end{figure}
     
\begin{figure}
    \centering    
    \begin{subfigure}[b]{0.49\textwidth}
        \includegraphics[width=\textwidth]{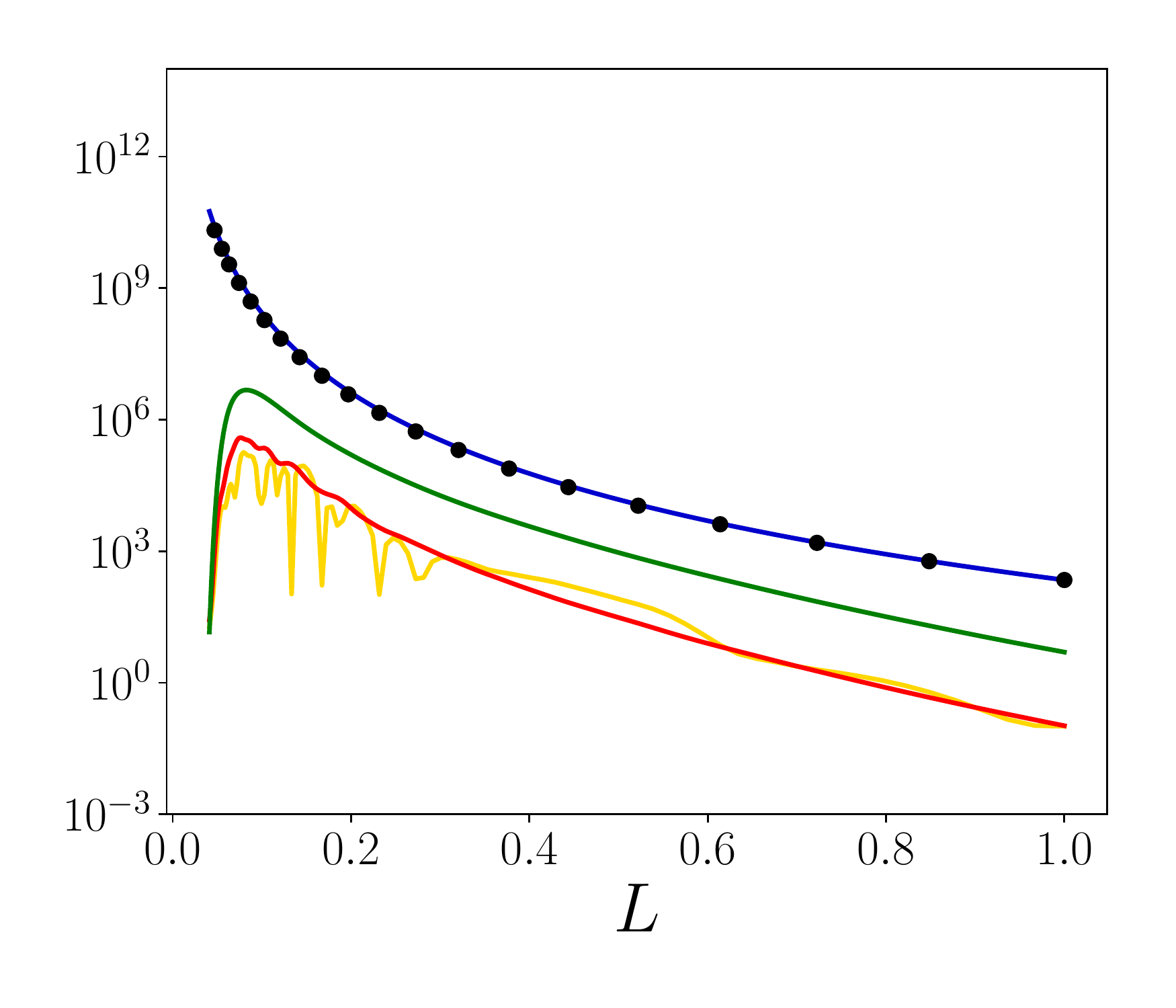}
        \label{fig:epsd_low_ie_bal_log_2}
    \end{subfigure}
    \begin{subfigure}[b]{0.49\textwidth}
        \includegraphics[width=\textwidth]{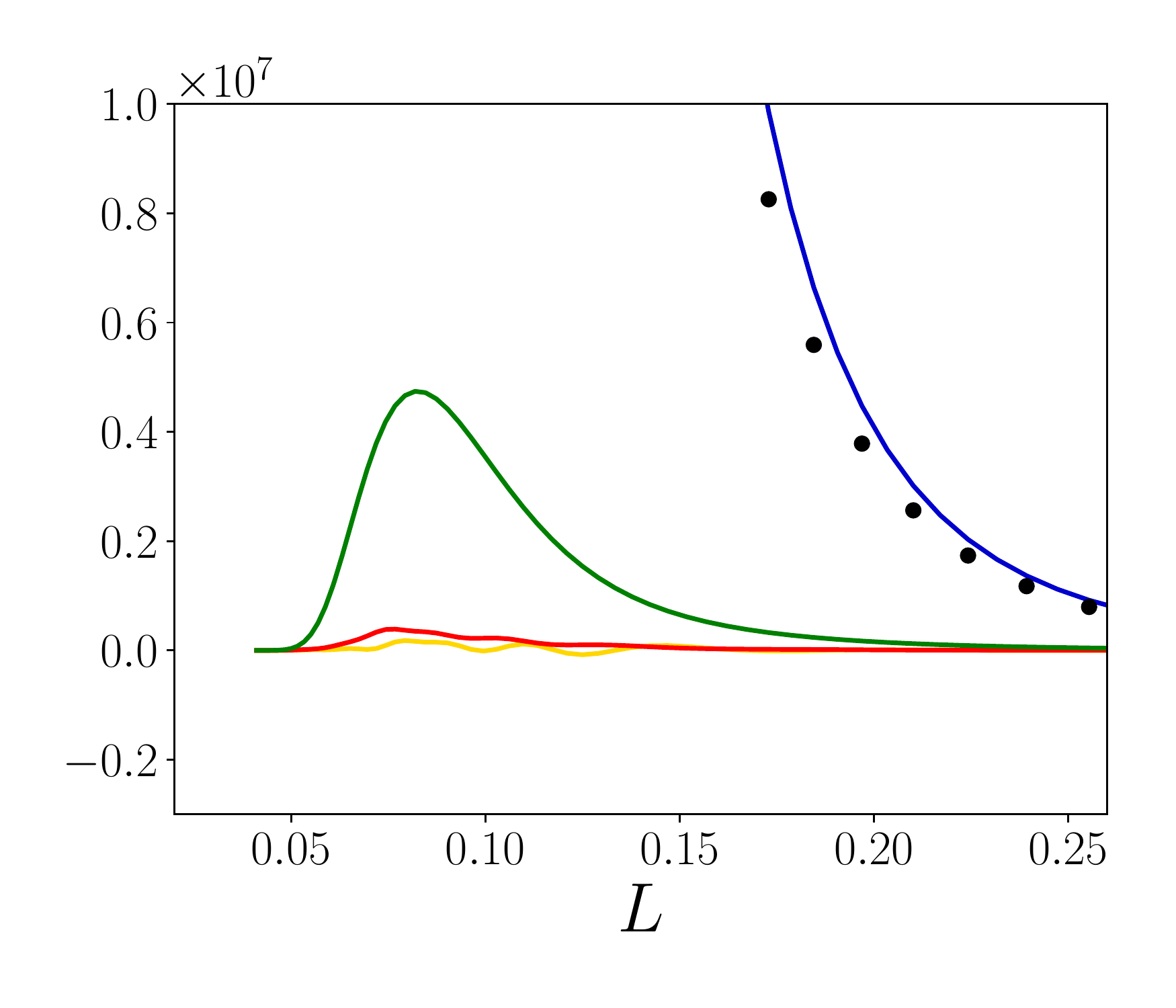}
        \label{fig:epsd_low_ie_bal_linear_2}
    \end{subfigure}     
    \caption{Mean internal energy budget in log scale on the left and linear scale on the right, for $S^* = 5.0$. The initial length of the domain is $L=1$, which decreases as time progresses. All terms have been normalized by $\rho_0 U_0^3 / L_0$. The absolute value of $PD$ is shown for the log-scale plot. The same legend as that of \cref{fig:epsd_low_ie_budget_1} applies to the plots above.}
    \label{fig:epsd_low_ie_budget_2}
\end{figure}

\begin{figure}
    \centering    
    \begin{subfigure}[b]{0.49\textwidth}
        \includegraphics[width=\textwidth]{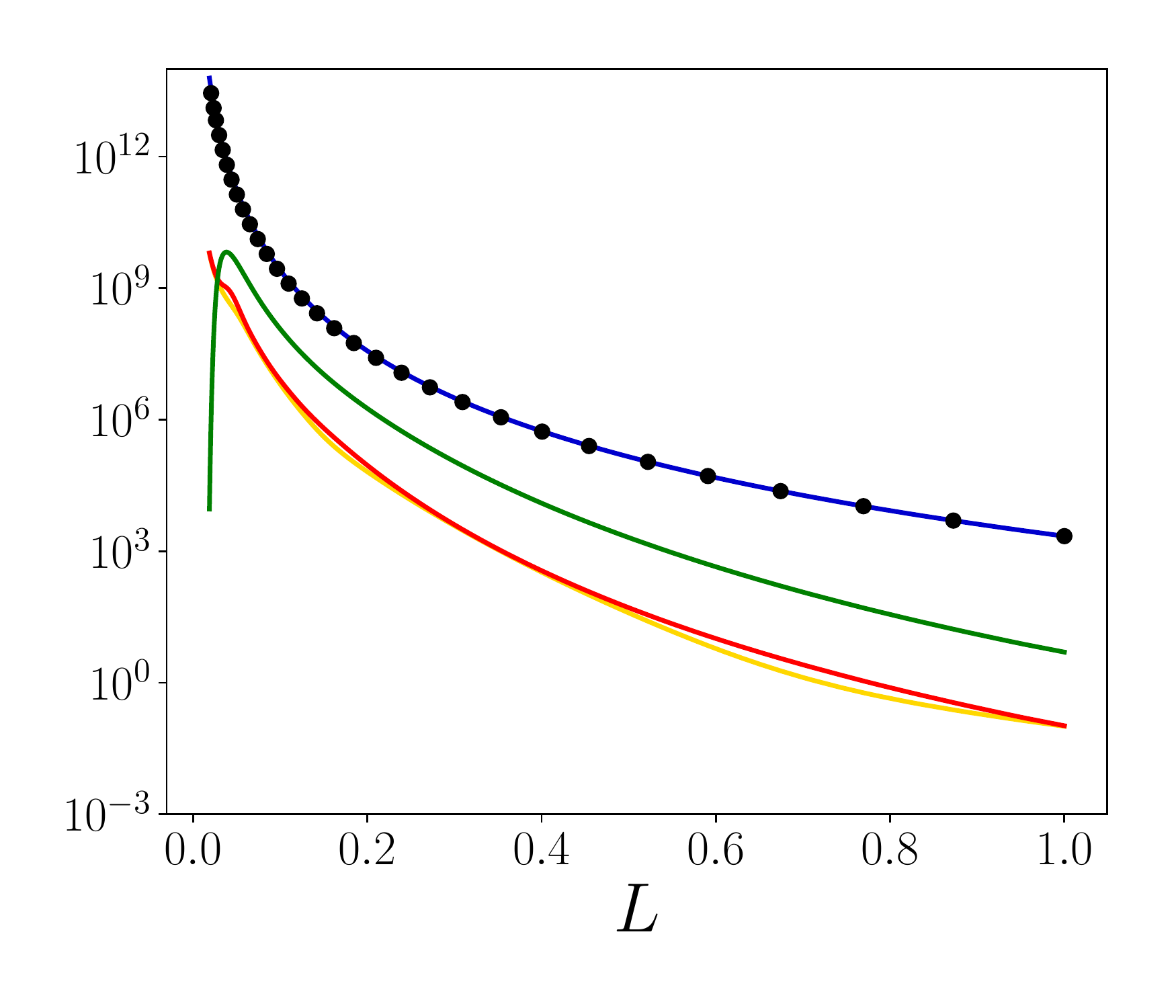}
        \label{fig:epsd_low_ie_bal_log_3}
    \end{subfigure}
    \begin{subfigure}[b]{0.49\textwidth}
        \includegraphics[width=\textwidth]{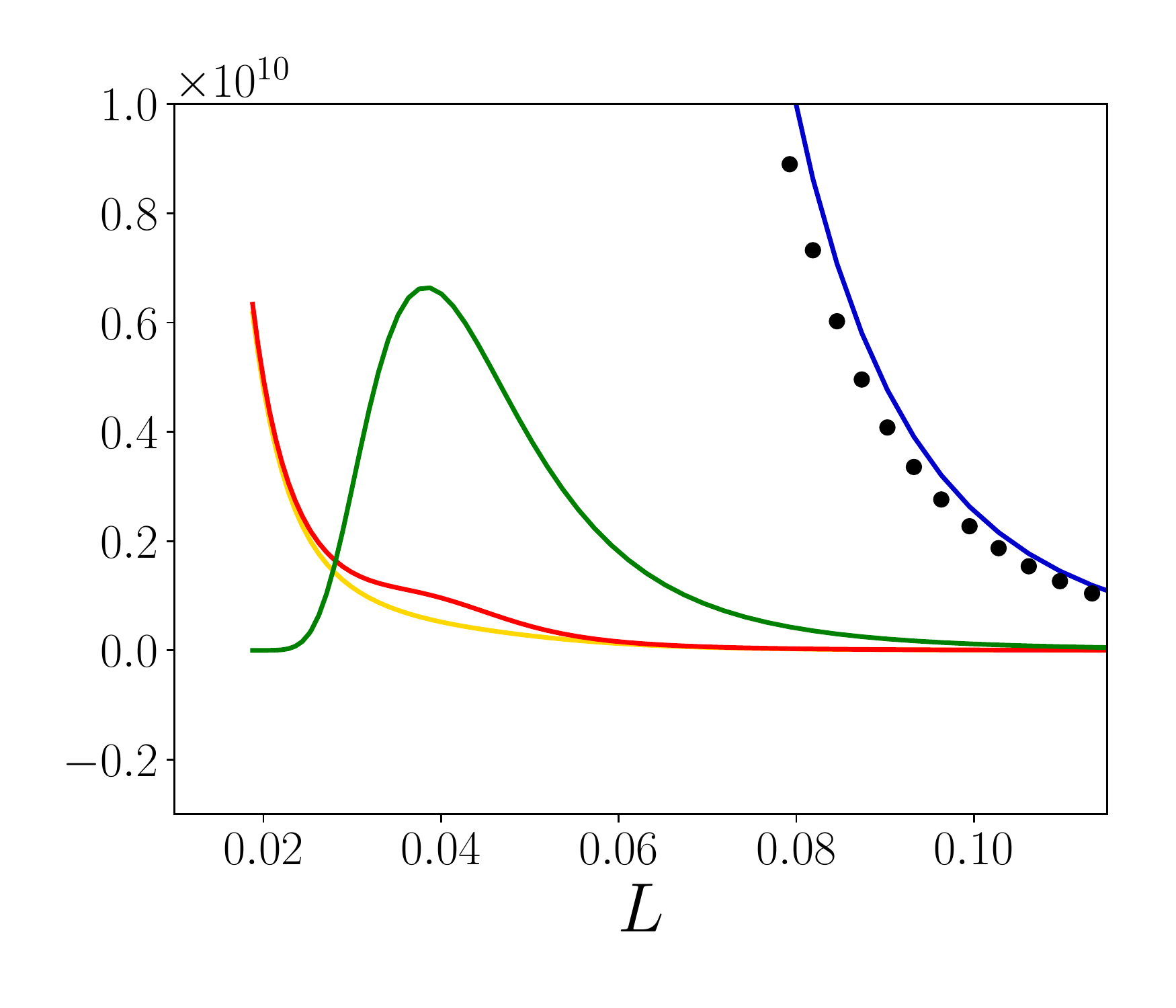}
        \label{fig:epsd_low_ie_bal_linear_3}
    \end{subfigure} 
    \caption{Mean internal energy budget in log scale on the left and linear scale on the right, for $S^* = 50$. The initial length of the domain is $L=1$, which decreases as time progresses. All terms have been normalized by $\rho_0 U_0^3 / L_0$. The absolute value of $PD$ is shown for the log-scale plot. The same legend as that of \cref{fig:epsd_low_ie_budget_1} applies to the plots above.}
    \label{fig:epsd_low_ie_budget_3}
\end{figure}       

\begin{figure}
    \centering    
    \begin{subfigure}[b]{0.49\textwidth}
        \includegraphics[width=\textwidth]{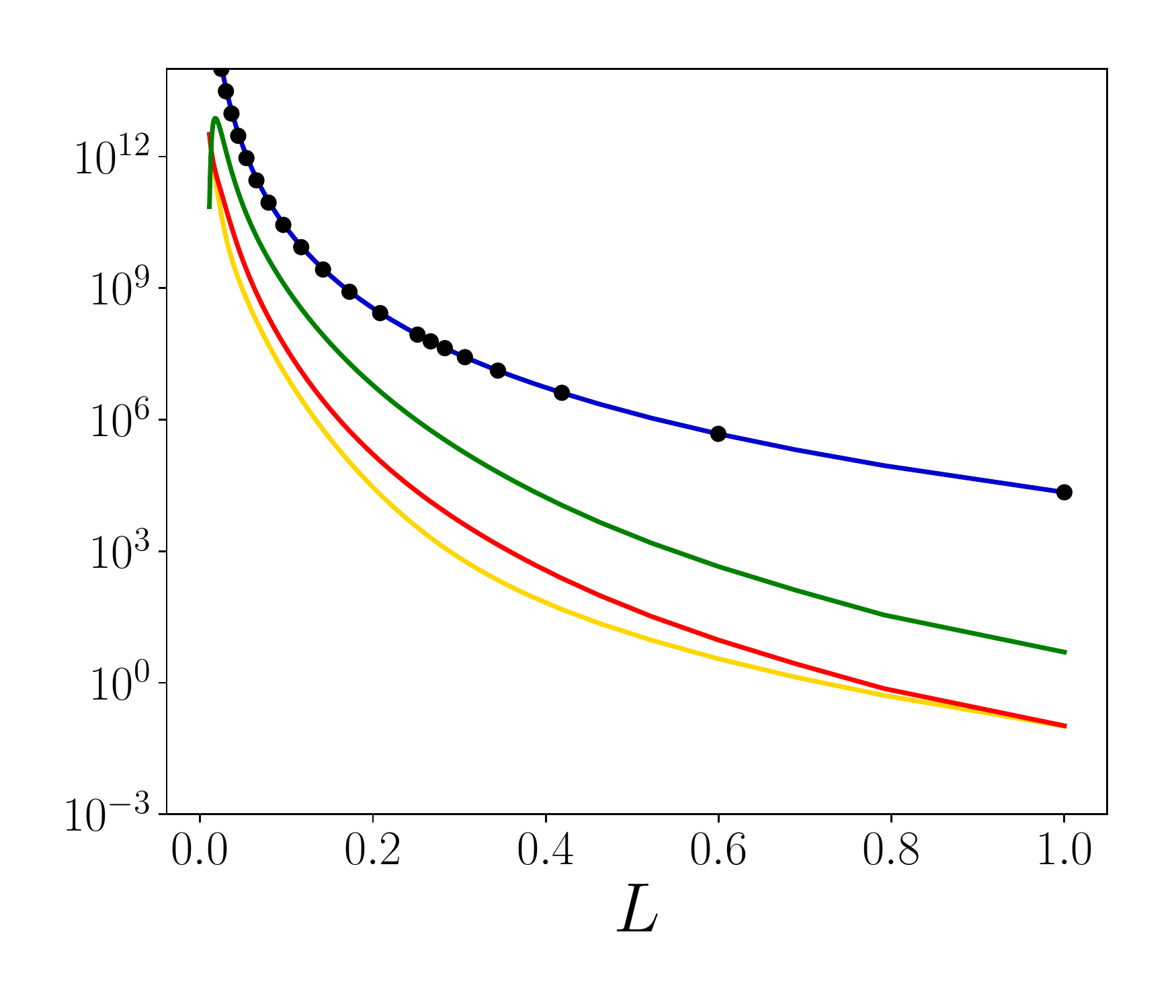}
        \label{fig:epsd_low_ie_bal_log_4}
    \end{subfigure}
    \begin{subfigure}[b]{0.49\textwidth}
        \includegraphics[width=\textwidth]{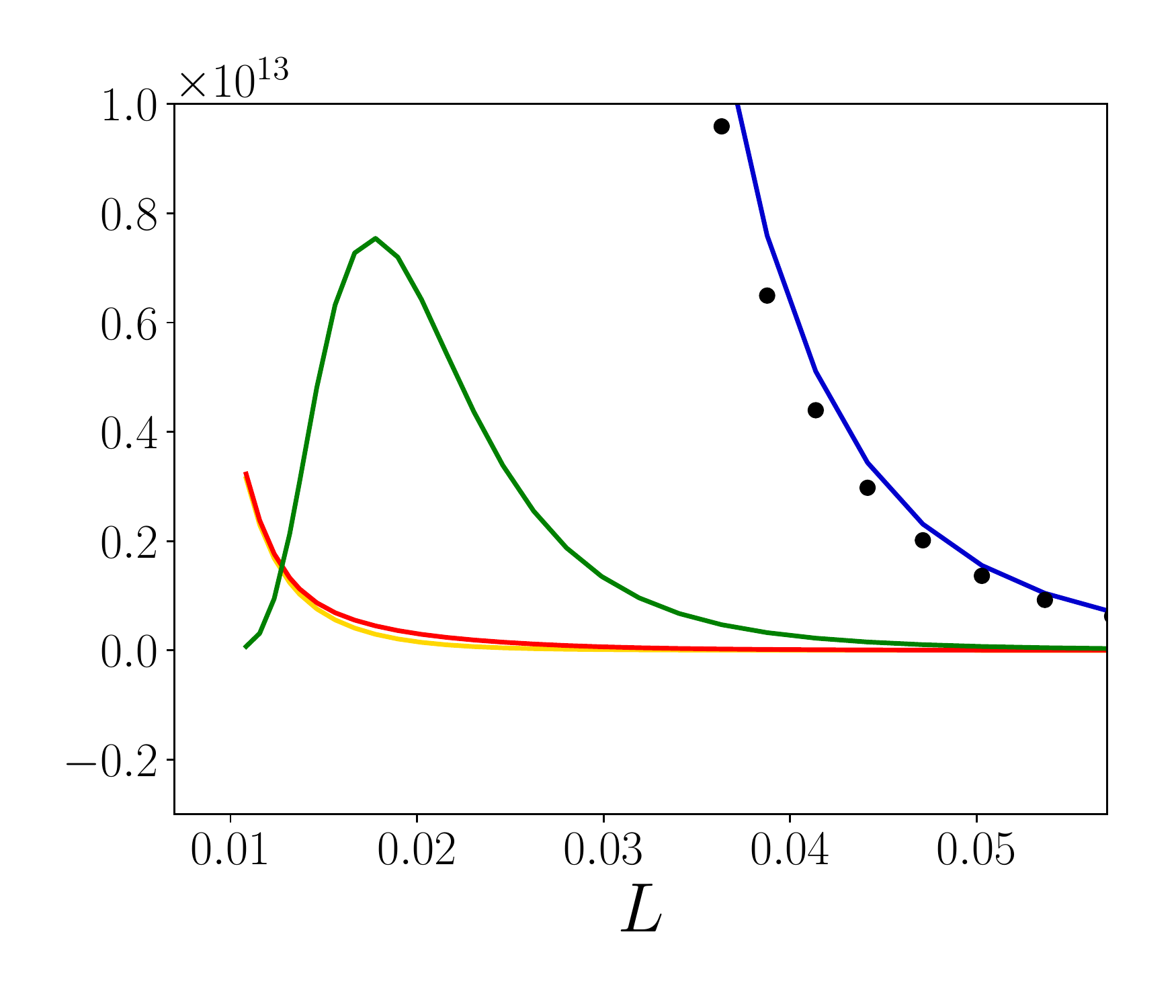}
        \label{fig:epsd_low_ie_bal_linear_4}
    \end{subfigure}    
    \caption{Mean internal energy budget in log scale on the left and linear scale on the right, for $S^* = 500$. The initial length of the domain is $L=1$, which decreases as time progresses. All terms have been normalized by $\rho_0 U_0^3 / L_0$. The absolute value of $PD$ is shown for the log-scale plot. The same legend as that of \cref{fig:epsd_low_ie_budget_1} applies to the plots above.}
    \label{fig:epsd_low_ie_budget_4}
\end{figure}

The temperature evolutions as a function of the domain length are shown in \cref{fig:epsd_low_T} for all the compression speeds. These are also compared against the $1/L^2$ temperature scaling corresponding to an adiabatic isentropic process with $\gamma = 5/3$, as assumed in \cite{davidovits2016}. As the figure shows, the temperature evolutions are in very close agreement with the adiabatic scaling. This indicates that the terms in the mean internal energy equation neglected under the assumption of adiabatic compression, namely the solenoidal dissipation, dilatational dissipation, and pressure dilatation, do not provide a strong contribution towards the increase of temperature for the current simulations.  

The negligible effect of the dissipations and the pressure dilatation is confirmed by comparing the source terms of the mean internal energy, as is done in \cref{fig:epsd_low_ie_budget_1,fig:epsd_low_ie_budget_2,fig:epsd_low_ie_budget_3,fig:epsd_low_ie_budget_4}. These figures show that, throughout the compression, the dominant source for the mean internal energy equation is the mechanical work, which takes the form of $MW =-3 \langle P \rangle \dot{L} / L$ for the given isotropic compression of \cref{eq:def_tensor}. For all compression speeds tested, the solenoidal dissipation, dilatational dissipation, and pressure dilatation are eclipsed by the mechanical work at all times during the compression. However, for the two fastest compression rates, the peak values of the dilatational dissipation and pressure dilatation are not achieved by the last-available simulated instance in time. Nonetheless, as shown in \cref{fig:epsd_low_tke}, by this last simulated instance in time the dilatational TKE has already surpassed its peak value and has dissipated by more than an order of magnitude, and it is thus unlikely that the dilatational dissipation and pressure dilatation will ever overtake the mechanical work. For the specific-heat ratio of $\gamma = 5/3$ and the assumption of an adiabatic compression, the mechanical work scales as $1/L^6$, which is shown as black dots in \cref{fig:epsd_low_ie_budget_1,fig:epsd_low_ie_budget_2,fig:epsd_low_ie_budget_3,fig:epsd_low_ie_budget_4}. Since the mechanical work overpowers the other sources of mean internal energy, it is expected that the assumption of an adiabatic compression would hold as well as shown in \cref{fig:epsd_low_ie_budget_1,fig:epsd_low_ie_budget_2,fig:epsd_low_ie_budget_3,fig:epsd_low_ie_budget_4}.

The dominance of the mechanical work can be further exemplified by considering the integrated values of the mean internal energy sources, shown in \Cref{tb:energy_integrated_sources}. The time-integrated contribution towards the increase of temperature due to mechanical work is at least three orders of magnitude larger than the second most significant time-integrated source term, namely, the solenoidal dissipation. An alternate metric for highlighting the dominance of the mechanical work is the comparison of the time-integrated contribution from the TKEs to mean internal energy against the time-integrated contribution from the mean kinetic energy to the mean internal energy. The ratio of these two factors for the four cases $S^*_0 = 0.50, 5.0, 50, \text{and, } 500$ is 0.0005, 0.0006, 0.0008, and 0.002, respectively.

Given that, for the parameters used in these simulations, the dissipated turbulent kinetic energy does not significantly increase the temperature of the system above the adiabatic prediction, it is crucial to determine under which conditions would the dissipated TKE actually lead to meaningful increases in temperature. To do this, we make use of the relation
\begin{equation}
\label{eq:cons_energy_ratio}
\frac{d}{dt} \left ( \frac{\widetilde{U} + k}{\widetilde{U}^{(a)}} \right) = 0,
\end{equation}
which was derived in \cref{sec:cons_energy_ratio}. $\widetilde{U}^{(a)}$ is the mean internal energy of the system given the idealized adiabatic compression, and is thus given by $\widetilde{U}^{(a)} = \widetilde{U}_0 L^{-2}$ where $\widetilde{U}_0$ is the initial value of $\widetilde{U}$. Integrating from the initial time $t_0$ to a final time $t_f$, one obtains
\begin{equation}
\label{eq:gain_factor_intermediate}
\left . \frac{\widetilde{U} + k}{\widetilde{U}^{(a)}} \right |_{t_f} = \left . \frac{ \widetilde{U} + k}{\widetilde{U}^{(a)} } \right |_{t_0} = 1 + \frac{k_0}{\widetilde{U}_0} = 1 + \frac{5}{9} M_{u,0}^2 .
\end{equation}
In the above we have made use of the definition of the fluctuating Mach number \cite{blaisdell1991}
\begin{equation}
M_u = \frac{\sqrt{ \widetilde{ u''_i u''_i}}}{ c(\widetilde{T})},
\end{equation}
whose initial value is denoted by $M_{u,0}$. We introduce $\widetilde{T}^{(a)} = \widetilde{T}_0 L^{-2}$ as the temperature corresponding to an adiabatic compression. If we define $t_f$ as the time by which all of the turbulent kinetic energy has dissipated, and $\widetilde{T}_f$ and $\widetilde{T}^{(a)}_f$ as the temperatures $\widetilde{T}$ and $\widetilde{T}^{(a)}$ at times $t > t_f$, respectively, then \cref{eq:gain_factor_intermediate} can be expressed as
\begin{equation}
\label{eq:gain_factor}
\frac{\widetilde{T}_f}{\widetilde{T}^{(a)}_f} = 1 + \frac{5}{9} M_{u,0}^2 .
\end{equation}
The above relation highlights a few notable aspects of the compression mechanism. Given $\widetilde{T}_0$ and $k_0$, $M_{u,0}$ is known, which, along with values of $L$ smaller than those corresponding to the time $t_f$, can be used in \cref{eq:gain_factor} to obtain the temperature achieved after the TKE has been fully dissipated. It is therefore unnecessary to carry out expensive numerical simulations to predict the temperature of the system post TKE depletion. The second aspect to highlight is that the temperature ratio $\widetilde{T}_f / \widetilde{T}^{(a)}_f$ is independent of the compression speed. However, the relationship in \cref{eq:gain_factor} holds for times $t > t_f$, where $t_f$ is different for the various compression speeds. 

For the simulations described in this paper, the initial fluctuating Mach number immediately preceding the start of the compression is $M_{u,0} = 0.651$. Using \cref{eq:gain_factor}, this gives $\widetilde{T}_f / \widetilde{T}^{(a)}_f = 1.235$. \Cref{tb:final_temp_ratio} lists this ratio computed from simulation data available at the last simulated instance in time, for the four compression speeds. As the table shows, there is strong agreement with the analytical value of 1.235. The slightly lower ratio for the fastest compression is most likely due to the fact that all of the TKE, specially the dilatational TKE, has not yet fully dissipated into heat. \Cref{eq:gain_factor} can now thus be used to predict under which conditions the dissipated TKE would lead to meaningful increases in temperature. For subsonic initial fluctuating Mach numbers, the temperature post TKE depletion can be up to about 1.5 times larger than that obtained with an adiabatic compression. If supersonic Mach numbers are used, such as $M_{u,0} = 2$ and 5, then the temperature post TKE depletion would be about 3 and 15 times larger, respectively, than for an adiabatic compression. For highly supersonic turbulence such as that encountered in the interstellar medium \cite{federrath2013,konstandin2016}, a Mach number of $M_{u,0} = 17 $ would lead to final temperatures about 160 times higher than those predicted assuming an adiabatic scaling. As stated in \cite{davidovits2018}, the hot spot of an inertial-confinement-fusion capsule can be characterized by a turbulent Mach number $M_t \approx 0.4$. Using this value in \cref{eq:gain_factor} leads to $\widetilde{T}_f/\widetilde{T}_f^{(a)} \approx 1.09$. This increase of temperature is minimal, and is eclipsed by the effect of the mechanical work. For example, if we assume that the sudden viscous dissipation of TKE occurs at $L = 0.1$, a small reduction of the domain size to $L = 0.0958$ would already allow the mechanical work to generate an equivalent increase in temperature. It is thus expected that only for flow fields with large initial Mach numbers would the self-consistent feedback mechanism lead to sudden dissipations with significant effects.

\begin{table}
\centering
\caption{Ratio $\widetilde{T}_f / \widetilde{T}^{(a)}_f$ obtained at the last simulated instance in time.}
\label{tb:final_temp_ratio}
\begin{tabular}{ Sc Sc Sc Sc Sc}
 & $S^*_0 = 0.50$ & $S^*_0 = 5.0$ & $S^*_0 = 50$ & $S^*_0 = 500$ \\
\hline\hline
$\frac{\widetilde{T}_f }{ \widetilde{T}^{(a)}_f } $ & 1.232 & 1.232 & 1.232 & 1.230 \\
\end{tabular}
\end{table}

\section{Concluding remarks}
\label{sec:conclusion}
A sudden viscous dissipation of plasma turbulence under compression was demonstrated in \cite{davidovits2016}. We expand on this previous work by accounting for the self-consistent feedback loop associated with this viscous mechanism. The feedback loop entails a transfer of energy from the turbulence towards the internal energy, and the subsequent increased temperatures and viscosities that in turn accelerate the original dissipation of TKE. Although previous efforts have reproduced the sudden dissipation of TKE, these do not capture the subsequent effect of the dissipated energy on the temperature, and the consequences thereof. This limitation is due to the use of the zero-Mach-limit assumption. To capture the increase of internal energy resulting from the dissipated TKE, and thus account for the entire self-consistent feedback loop, direct numerical simulations have been carried out using a finite-Mach number formulation that solves transport equations for the density, fluctuating velocity, and total energy. The analysis of the self-consistent feedback loop was divided into two steps: the first focused on the evolution of the solenoidal and dilatational TKEs, and the second on the evolution of the mean internal energy as it absorbs the dissipated TKE.

Results show that both the solenoidal and dilatational TKE exhibit the sudden viscous dissipation mechanism, with the dissipation of dilatational TKE slightly lagging that of solenoidal TKE. Moreover, large oscillations in the temporal evolution of dilatational TKE for slow compression rates are observed, which are correlated with the highly-oscillatory nature of the pressure dilatation. The pressure dilatation constitutes a two-way energy transfer between the dilatational TKE and the mean internal energy of the system. However, a detailed analysis of the dilatational TKE budget shows that the time-integrated effect of the pressure dilatation is to transfer energy from heat towards dilatational TKE, even for cases when the pressure dilatation transfers energy in both directions on short time scales. Thus, the dilatational dissipation needs to overcome both the dilatational production and pressure dilatation for the sudden viscous dissipation of dilatational TKE to take place, which explains the delayed dissipation of dilatational TKE with respect to solenoidal TKE.

An analysis of the sources for the mean internal energy shows that mechanical work, which transforms energy from the mean flow to increase heat, dominates all other sources of mean internal energy for the turbulent Mach numbers chosen in this study. For all instances in time, the mechanical work term is larger, often by multiple orders of magnitude, than the solenoidal and dilatational dissipation and the pressure dilatation. As a result, the contribution of the dissipated TKE towards the increase of temperature is minimal, and the temperature evolution closely follows an adiabatic scaling. This validates previous efforts \cite{davidovits2016,davidovits2016b,davidovits2017,viciconte2018} that relied on a fixed adiabatic scaling for the temperature evolution. An analytical expression was also derived for the ratio of the temperature post TKE depletion to the temperature obtained at the same instance of the compression assuming an adiabatic scaling. This ratio depends on the initial fluctuating Mach number only, indicating that the rate of compression does not affect the magnitudes of the temperature post TKE depletion. The derived analytical expression also shows that for subsonic initial fluctuating Mach numbers, the true temperature of the system is not substantially larger than that obtained if the TKE contribution is neglected. To provide a point of reference, it was shown that for an adiabatic compression where only the mechanical work is active, reducing the domain from $L = 0.1$ to $L = 0.0958$ would have an equivalent effect on temperature as that of the suddenly dissipated TKE. This thus indicates that the potential of the sudden viscous dissipation mechanism to significantly enhance the heating of the plasma by dissipating the inherent turbulence could be limited. Nonetheless, this mechanism could serve as an effective tool to prevent the occurrence of turbulent fluctuations, which are responsible for detrimental mixing in inertial-confinement-fusion capsules. It is crucial to highlight that the finite-Mach-number framework chosen here, although more general than the zero-Mach-number formalism, is still missing physics relevant to inertial-confinement fusion such as non-ideal equations of state, radiation transport, multiple species, plasma viscosity models, separate ion and electron temperatures, and alpha heating. Thus, these factors need to be explored to provide a definite assessment on the ability of the sudden viscous dissipation mechanism to improve the performance of inertial confinement fusion.  

\section{Acknowledgments}
The authors wish to thank Dr.\@ O.\@ Schilling for helpful comments on drafts of the manuscript. This work was performed under the auspices of the U.S. Department of Energy by Lawrence Livermore National Laboratory under Contract DE-AC52-07NA27344. 

\appendix

\section{Derivation of the finite-Mach-number Navier-Stokes equations for isotropic mean compression}
\label{sec:comp_fluc_NS_mean_comp}
The derivation of the governing equations used for the computational simulations in this study is detailed below. This derivation is divided into five distinct steps, each described in the five subsections below.

\subsection{Compressible Navier-Stokes equations}
The starting point are the Navier-Stokes equations for a compressible fluid. Thus, the evolution of the density $\rho =\rho(t,\bold{x})$, velocity $U_i =U_i(t,\bold{x})$ and total energy $E=E(t,\bold{x})$ is governed by
\begin{equation}
\label{eq:rho}
\frac{\partial \rho}{\partial t} + \frac{ \partial \rho U_i}{ \partial x_i} = 0,
\end{equation}
\begin{equation}
\label{eq:rhou}
\frac{\partial \rho U_i}{\partial t}+\frac{\partial \rho U_i U_j}{\partial x_j} = \frac{\partial \sigma_{ij}}{\partial x_j} ,
\end{equation}
\begin{equation}
\label{eq:rhoE}
\frac{\partial \rho E}{\partial t} + \frac{\partial \rho E U_j}{\partial x_j} = \frac{\partial U_i \sigma_{ij}}{\partial x_j} + \frac{\partial}{\partial x_j} \left ( \kappa \frac{\partial T}{\partial x_j} \right ).
\end{equation}
Closure of the above is achieved with 
\begin{equation}
\label{eq:sigma}
    \sigma_{ij} = -P \delta_{ij} + 2\mu \left[ \frac{1}{2} \left( \frac{\partial U_i}{\partial x_j} + \frac{\partial U_j}{\partial x_i} \right ) - \frac{1}{3} \frac{\partial U_k}{\partial x_k} \delta_{ij} \right] ,
\end{equation}
\begin{equation}
     E = U + K ,
\end{equation}
\begin{equation} 
    U = C_v T \qquad K = \frac{1}{2} U_i U_i ,
\end{equation}
\begin{equation}
    P = \rho R T ,
\end{equation}
\begin{equation}
    \kappa = \frac{\mu C_p}{Pr} ,
\end{equation}
\begin{equation}
    \mu = \mu_0 \left(\frac{T}{T_{0}}\right)^{n}. 
\end{equation}
$P=P(t,\bold{x})$ is the pressure, $T=T(t,\bold{x})$ the temperature, $U=U(t,\bold{x})$ the internal energy, $K=K(t,\bold{x})$ the kinetic energy, $\mu = \mu(t,\bold{x})$ the dynamic viscosity, and $\kappa =\kappa(t,\bold{x})$ the thermal conductivity. $C_v$, $C_p$, $R$, and $Pr$ are the specific heat at constant volume, the specific heat at constant pressure, the ideal gas constant, and the Prandtl number, respectively. For the power law of viscosity, $\mu_0$ and $T_0$ represent reference viscosity and temperature values, and $n$ is the power-law exponent.

\subsection{Homogeneous turbulence}
We summarize here and in the following subsection the derivations carried out by \cite{blaisdell1991} to obtain the governing equations for homogeneous compressible turbulence. The quantities $\langle \rho \rangle$ and $\langle P \rangle$ are defined as Reynolds-averaged density and pressure, respectively, and $\widetilde{U}_i$ as the Favre-averaged velocity.
\cite{blaisdell1991} showed that for turbulence to remain homogeneous, necessary and sufficient conditions are that $\langle \rho \rangle $ and $\langle P \rangle$ depend on $t$ but not $\bold{x}$, and that $\tilde{U}_i$ be given by 
\begin{equation}
\tilde{U}_i = G_{ij} x_j,
\end{equation}
where $G_{ij} = \frac{\partial \tilde{U}_i }{\partial x_j}$ also depends only on $t$ and not $\bold{x}$. Given the above assumptions, averaging of the momentum equation shows that the evolution of $G_{ij}$ is dictated by
\begin{equation}
\frac{dG_{ij}}{dt} + G_{kj}G_{ik} = 0.
\end{equation}
Moreover, using the assumptions above and plugging in the decomposition $U_i = \widetilde{U}_i + u''_i$ in \cref{eq:rho,eq:rhou,eq:rhoE,eq:sigma}, \cite{blaisdell1991} derived the governing equations in terms of the fluctuating velocity. These are
\begin{equation}
\label{eq:rho_f_G}
    \frac{\partial \rho}{\partial t} + \frac{\partial \rho}{\partial x_i} G_{ij} x_j + \frac{\partial \rho u''_i}{\partial x_i} = f^{(\rho)},
\end{equation}
\begin{equation}
\label{eq:rhou_f_G}
    \frac{\partial \rho u''_i}{\partial t} + \frac{\partial \rho u''_i}{\partial x_j} G_{jk} x_k + \frac{\partial \rho u''_i u''_j}{\partial x_j} = \frac{\partial \sigma_{ij}}{\partial x_j} + f^{(u)}_i ,
\end{equation}
\begin{equation}
\label{eq:rhoE_f_G}
    \frac{\partial \rho E_t}{\partial t} + \frac{\partial \rho E_t}{\partial x_i} G_{ik} x_k + \frac{\partial \rho E_t u''_i}{\partial x_i} = \frac{\partial u''_i \sigma_{ij}}{\partial x_j} + \frac{\partial}{\partial x_j} \left ( \kappa \frac{ \partial T}{\partial x_j} \right ) +f^{(e)}.
\end{equation}
Closure of the above is achieved with
\begin{equation}
    \sigma_{ij} = -P \delta_{ij} + 2\mu \left[ \frac{1}{2} \left( \frac{\partial u''_i}{\partial x_j} + \frac{\partial u''_j}{\partial x_i} \right ) - \frac{1}{3} \frac{\partial u''_k}{\partial x_k} \delta_{ij} \right] + 2\mu \left[  \frac{1}{2} \left ( G_{ij} + G_{ji} \right ) - \frac{1}{3} G_{ii} \delta_{ij} \right] ,
\end{equation}
\begin{equation}
    E_t = U + K_t ,
\end{equation}
\begin{equation}
    U = C_v T \qquad K_t = \frac{1}{2} u''_i u''_i ,
\end{equation}
\begin{equation}
    P = \rho R T ,
\end{equation}
\begin{equation}
    \kappa = \frac{\mu C_p}{Pr} , 
\end{equation}
\begin{equation}
    \mu = \mu_0 \left(\frac{T}{T_{0}}\right)^n , 
\end{equation}
\begin{equation}
\label{eq:frho}
    f^{(\rho)} = -\rho G_{ii} ,
\end{equation}
\begin{equation}
\label{eq:frhou}
    f^{(u)}_i = -\rho u''_j G_{ij} - \rho u''_i G_{jj} ,
\end{equation}
\begin{equation}
\label{eq:frhoe}
    f^{(e)} = - \rho E_t G_{ii} - \rho u''_i u''_j G_{ij} + G_{ij} \sigma_{ij} .
\end{equation}

\subsection{Rogallo transformation}
As is typically done for simulations of homogeneous turbulence (see for example \cite{rogallo1981,blaisdell1991}) one can reformulate the equations using a deforming reference frame---referred to here as the Rogallo reference frame---to eliminate those terms in \cref{eq:rho_f_G,eq:rhou_f_G,eq:rhoE_f_G} that have an explicit dependence on position. The variables in the Rogallo reference frame are denoted as $\rhorog = \rhorog(t,\xrogvec)$, $\urogf = \urogf(t,\xrogvec)$, $\Prog = \Prog(t,\xrogvec)$, $\Trog = \Trog(t,\xrogvec)$. The relationship between the variables in the original reference frame and the Rogallo reference frame is as
\begin{align}
    \rho &= \rhorog(t,\bold{f}) \nonumber \\
    u''_i & = \urogf_i(t,\bold{f}) \nonumber \\
    P & = \Prog(t,\bold{f}) \nonumber \\
    T & = \Trog(t, \bold{f}),
\end{align}
where $f_i = A_{ij} x_j$. $A_{ij}$ is referred to as the coordinate-transformation tensor, it depends on $t$ only, and is defined so as to satisfy
\begin{equation}
\frac{dA_{ij}}{dt} + A_{ik} G_{kj} = 0.
\end{equation}
Using this transformation, the governing equations in the Rogallo reference frame are
\begin{equation}
\label{eq:rho_A}
    \frac{\partial \rhorog}{\partial t} + \frac{\partial \rhorog \urogf_i}{\partial \xrog_j}A_{ji} = \mathring{f}^{(\rho)} ,
\end{equation}
\begin{equation}
\label{eq:rhou_A}
    \frac{\partial \rhorog \urogf_i}{\partial t} + \frac{\partial \rhorog \urogf_i \urogf_j}{\partial \xrog_k}A_{kj} = \frac{\partial \mathring{\sigma}_{ij}}{\partial \xrog_k} A_{kj} + \mathring{f}^{(u)}_i ,
\end{equation}
\begin{equation}
\label{eq:rhoE_A}
    \frac{\partial \rhorog \mathring{E}_t}{\partial t} + \frac{\partial \rhorog \mathring{E}_t \urogf_i}{\partial \xrog_j}A_{ji} = \frac{\partial \urogf_i \mathring{\sigma}_{ij}}{\partial \xrog_k}A_{kj} + \frac{\partial}{\partial \xrog_l} \left ( \mathring{\kappa} \frac{ \partial \Trog}{\partial \xrog_k} \right )A_{kj}A_{lj} + \mathring{f}^{(e)} .
\end{equation}
Closure of the above is achieved with
\begin{equation}
\label{eq:sigma_A}
    \mathring{\sigma}_{ij} = -\Prog \delta_{ij} + 2 \mathring{\mu} \left[ \frac{1}{2} \left( \frac{\partial \urogf_i}{\partial \xrog_n}A_{nj} + \frac{\partial \urogf_j}{\partial \xrog_n}A_{ni} \right ) - \frac{1}{3} \frac{\partial \urogf_k}{\partial \xrog_n}A_{nk} \delta_{ij} \right] + 2 \mathring{\mu} \left[  + \frac{1}{2} \left ( G_{ij} + G_{ji} \right ) - \frac{1}{3} G_{ii} \delta_{ij} \right] ,
\end{equation}
\begin{equation}
     \mathring{E}_t = \mathring{U} + \mathring{K}_t ,
\end{equation}
\begin{equation} 
    \mathring{U} = C_v \Trog \qquad \mathring{K}_t = \frac{1}{2} \urogf_i \urogf_i ,
\end{equation}
\begin{equation}
    \Prog = \rhorog R \Trog ,
\end{equation}
\begin{equation}
    \mathring{\kappa} = \frac{ \mathring{\mu} C_p}{Pr} ,
\end{equation}
\begin{equation}
    \mathring{\mu} = \mu_0 \left( \frac{\Trog}{T_0}\right)^n ,
\end{equation}
\begin{equation}
    \mathring{f}^{(\rho)} = -\rhorog G_{ii} ,
\end{equation}
\begin{equation}
    \mathring{f}^{(u)}_i = -\rhorog \urogf_j G_{ij} - \rhorog \urogf_i G_{jj} ,
\end{equation}
\begin{equation}
    \mathring{f}^{(e)} = - \rhorog \mathring{E}_t G_{ii} - \rhorog \urogf_i \urogf_j G_{ij} + G_{ij} \mathring{\sigma}_{ij} .
\end{equation}

\subsection{Isotropic compression}
The mean flow deformation for isotropic compression is given in \cite{wu1985,blaisdell1991,davidovits2016}, and can be expressed as
\begin{equation}
G_{ij} = \frac{\dot{L}}{L} \delta_{ij} ,
\end{equation}
where $\dot{L}$ is constant and thus $L = 1 + \dot{L} t$. The corresponding coordinate transformation tensor is 
\begin{equation}
A_{ij} = \frac{1}{L} \delta_{ij}.
\end{equation}
Thus, using the above in \cref{eq:rho_A,eq:rhou_A,eq:rhoE_A}, we obtain
\begin{equation}
    \frac{\partial \rhorog}{\partial t} + \frac{\partial \rhorog \urogf_i}{\partial \xrog_i}\frac{1}{L} = \mathring{f}^{(\rho)} ,
\end{equation}
\begin{equation}
    \frac{\partial \rhorog \urogf_i}{\partial t} + \frac{\partial \rhorog \urogf_i \urogf_j}{\partial \xrog_j} \frac{1}{L} = \frac{\partial \mathring{\sigma}_{ij}}{\partial \xrog_j} \frac{1}{L} + \mathring{f}_i^{(u)} ,
\end{equation}
\begin{equation}
    \frac{\partial \rhorog \mathring{E}_t}{\partial t} + \frac{\partial \rhorog \mathring{E}_t \urogf_i}{\partial \xrog_i} \frac{1}{L} = \frac{\partial \urogf_i \mathring{\sigma}_{ij}}{\partial \xrog_j} \frac{1}{L} + \frac{\partial}{\partial \xrog_j} \left ( \mathring{\kappa} \frac{ \partial \Trog}{\partial \xrog_j} \right ) \frac{1}{L^2} + \mathring{f}^{(e)}.
\end{equation}
Closure of the above is achieved with
\begin{equation}
    \mathring{\sigma}_{ij} = -\Prog \delta_{ij} + 2 \mathring{\mu} \left[ \frac{1}{2} \left( \frac{\partial \urogf_i}{\partial \xrog_j} \frac{1}{L} + \frac{\partial \urogf_j}{\partial \xrog_i}\frac{1}{L} \right ) - \frac{1}{3} \frac{\partial \urogf_k}{\partial \xrog_k}\frac{1}{L} \delta_{ij} \right] ,
\end{equation}
\begin{equation}
     \mathring{E}_t = \mathring{U} + \mathring{K}_t ,
\end{equation}
\begin{equation} 
    \mathring{U} = C_v \Trog \qquad \mathring{K}_t = \frac{1}{2} \urogf_i \urogf_i ,
\end{equation}
\begin{equation}
    \Prog = \rhorog R \Trog ,
\end{equation}
\begin{equation}
    \mathring{\kappa} = \frac{ \mathring{\mu} C_p}{Pr} ,
\end{equation}
\begin{equation}
    \mu = \mu_0 \left( \frac{\Trog}{T_0}\right)^n ,
\end{equation}
\begin{equation}
\mathring{f}^{(\rho)} = - 3 \rhorog \frac{\dot{L}}{L} ,
\end{equation}
\begin{equation}
\mathring{f}^{(u)}_i = -4 \rhorog \urogf_i \frac{\dot{L}}{L} ,
\end{equation}
\begin{equation}
\mathring{f}^{(e)} = - 3 \rhorog \mathring{E}_t \frac{ \dot{L} }{L} - \rhorog \urogf_i \urogf_i \frac{\dot{L}}{L} - 3 \Prog \frac{\dot{L}}{L}.
\end{equation}

\subsection{Re-scaling}
An additional transformation can be performed so that, as the simulation advances in time, division by very small values of $L$ is avoided. The analogue of this re-scaling for the zero-Mach limit is detailed in \cite{davidovits2016} and in the Appendix of \cite{davidovits2016b}. The new re-scaled flow variables are $\hat{\rho} = \hat{\rho}(\hat{t},\xrogvec)$, $\hat{u}''_i = \hat{u}''_i(\hat{t},\xrogvec)$, $\hat{P} = \hat{P}(\hat{t},\xrogvec)$, and $\hat{T} = \hat{T}(\hat{t},\xrogvec)$. Their relation to the original variables is
\begin{align}
\rhorog &= \hat{\rho}(g,\xrogvec) L^{-1} \nonumber \\
\urogf_i &= \hat{u}''_i(g,\xrogvec) \nonumber \\
\Prog &= \hat{P}(g,\xrogvec) L^{-1} \nonumber \\
\Trog &= \hat{T}(g,\xrogvec),
\end{align}
where $g = g(t)$ is defined by $\frac{dg}{dt} = L^{-1}$. We define $\hat{\mu} = \hat{\mu}(\hat{t},\xrogvec)$ and $\hat{\kappa} = \hat{\kappa}(\hat{t},\xrogvec)$ as
\begin{equation}
\hat{\mu} = \mu_0 \left ( \frac{\hat{T}}{T_0} \right)^n
\end{equation}
and
\begin{equation}
\hat{\kappa} = \frac{\hat{\mu} C_p}{Pr}
\end{equation}
so that $\mathring{\mu} = \hat{\mu}(g,\xrogvec)$ and $\mathring{\kappa} = \hat{\kappa}(g,\xrogvec)$. The re-scaled stress tensor $\hat{\sigma} = \hat{\sigma}(\hat{t},\xrogvec)$ is defined as
\begin{equation}
\hat{\sigma}_{ij} = -\hat{P} \delta_{ij} + 2\hat{\mu} \left [ \frac{1}{2} \left ( \frac{\partial \hat{u}''_i}{\partial \xrog_j} + \frac{\partial \hat{u}''_j}{\partial \xrog_i} \right ) - \frac{1}{3} \frac{\partial \hat{u}''_k}{\partial \xrog_k} \delta_{ij} \right ] ,
\end{equation}
so that $\mathring{\sigma}_{ij} = \hat{\sigma}_{ij}(g,\xrog) L^{-1}$.

Using the re-scaled variables above, the continuity equation transforms as follows
\begin{equation*}
\left (\frac{\partial \hat{\rho}}{\partial \hat{t}} \right)_{\hat{t} = g} \frac{1}{L^2} - ( \hat{\rho} )_{\hat{t} = g} \frac{ \dot{L}}{L^2} + \left ( \frac{\partial \hat{\rho} \hat{u}''_i}{\partial \xrog_i} \right)_{\hat{t} = g} \frac{1}{L^2} = -3 ( \hat{\rho} )_{\hat{t} = g} \frac{\dot{L}}{L^2}.
\end{equation*}
The conservation of momentum equation becomes
\begin{equation*}
\left (\frac{\partial \hat{\rho} \hat{u}''_i}{\partial \hat{t}} \right )_{\hat{t} = g}\frac{1}{L^2} - ( \hat{\rho} \hat{u}''_i)_{\hat{t} = g} \frac{\dot{L}}{L^2} + \left ( \frac{\partial \hat{\rho} \hat{u}''_i \hat{u}''_j}{\partial \xrog_j} \right)_{\hat{t} = g} \frac{1}{L^2} = \left ( \frac{\partial \hat{\sigma}_{ij}}{\partial \xrog_j} \right)_{\hat{t} = g} \frac{1}{L^2} -4 ( \hat{\rho} \hat{u}''_i)_{\hat{t} = g} \frac{\dot{L}}{L^2}.
\end{equation*}
The total energy $\hat{E}_t = \hat{E}_t(\hat{t},\xrogvec)$ is defined as $\hat{E}_t = \hat{U} + \hat{K}_t$, where $\hat{U} = C_v \hat{T}$ and $\hat{K}_t = \frac{1}{2} \hat{u}''_i \hat{u}''_i$, so that $\mathring{E}_t = \hat{E}_t(g,\xrogvec)$. The energy equation can thus be expressed as
\begin{multline*}
\left ( \frac{ \partial \hat{\rho} \hat{E}_t }{\partial \hat{t}} \right)_{\hat{t} = g} \frac{1}{L^2} - ( \hat{\rho} \hat{E}_t )_{\hat{t} = g} \frac{\dot{L}}{L^2} + \left ( \frac{\partial \hat{\rho} \hat{E}_t \hat{u}''_i}{\partial \xrog_i} \right)_{\hat{t} = g} \frac{1}{L^2} = \\
 \left ( \frac{\partial \hat{u}''_i \hat{\sigma}_{ij} }{\partial x_j} \right )_{\hat{t} = g} \frac{1}{L^2} + \left [ \frac{\partial}{\partial x_j} \left ( \hat{\kappa} \frac{\partial \hat{T}}{\partial x_j} \right ) \right ]_{\hat{t} = g} \frac{1}{L^2} - 3 (\hat{\rho} \hat{E}_t )_{\hat{t} = g} \frac{\dot{L}}{L^2} - (\hat{\rho} \hat{u}''_i \hat{u}''_i)_{\hat{t} = g} \frac{\dot{L}}{L^2} - 3 (\hat{P})_{\hat{t} = g} \frac{\dot{L}}{L^2}.
\end{multline*}
The $1/L^2$ factors can now be eliminated. Evaluating the equations at $t = g^{-1}(\hat{t})$ we finally obtain
\begin{equation*}
\frac{\partial \hat{\rho}}{\partial \hat{t}} + \frac{\partial \hat{\rho} \hat{u}''_i}{\partial \xrog_i} = -2 \hat{\rho} \left ( \dot{L} \right )_{t = g^{-1}(\hat{t})} ,
\end{equation*}
\begin{equation*}
\frac{\partial \hat{\rho} \hat{u}''_i}{\partial \hat{t}} + \frac{\partial \hat{\rho} \hat{u}''_i \hat{u}''_j}{\partial \xrog_j} = \frac{\partial \hat{\sigma}_{ij}}{\partial \xrog_j} - 3 \hat{\rho} \hat{u}''_i \left ( \dot{L} \right )_{t = g^{-1}(\hat{t})} ,
\end{equation*}
\begin{equation*}
\frac{\partial \hat{\rho} \hat{E}_t}{\partial \hat{t}} + \frac{\partial \hat{\rho} \hat{E}_t \hat{u}''_i}{\partial \xrog_i} = \frac{\partial \hat{u}''_i \hat{\sigma}_{ij}}{\partial \xrog_j} + \frac{\partial}{\partial \xrog_j} \left ( \hat{\kappa} \frac{ \partial \hat{T}}{\partial \xrog_j} \right ) - \left ( 2 \hat{\rho} \hat{E}_t  + \hat{\rho} \hat{u}''_i \hat{u}''_i + 3 \hat{P}\right) \left ( \dot{L} \right )_{t = g^{-1}(\hat{t})} .
\end{equation*}
Tu summarize, we write the governing equations as
\begin{equation}
\label{eq:rho_modframe}
\frac{\partial \hat{\rho}}{\partial \hat{t}} + \frac{\partial \hat{\rho} \hat{u}''_i}{\partial \xrog_i} = \hat{f}^{(\rho)} ,
\end{equation}
\begin{equation}
\label{eq:rhou_modframe}
\frac{\partial \hat{\rho} \hat{u}''_i}{\partial \hat{t}} + \frac{\partial \hat{\rho} \hat{u}''_i \hat{u}''_j}{\partial \xrog_j} = \frac{\partial \hat{\sigma}_{ij}}{\partial \xrog_j} + \hat{f}_i^{(u)} ,
\end{equation}
\begin{equation}
\label{eq:rhoE_modframe}
\frac{\partial \hat{\rho} \hat{E}_t}{\partial \hat{t}} + \frac{\partial \hat{\rho} \hat{E}_t \hat{u}''_i}{\partial \xrog_i} = \frac{\partial \hat{u}''_i \hat{\sigma}_{ij}}{\partial \xrog_j} + \frac{\partial}{\partial \xrog_j} \left ( \hat{\kappa} \frac{ \partial \hat{T}}{\partial \xrog_j} \right ) + \hat{f}^{(e)} .
\end{equation}
Closure of the above is achieved with
\begin{equation}
\label{eq:sigma_modframe}
\hat{\sigma}_{ij} = -\hat{P} \delta_{ij} + 2\hat{\mu} \left [ \frac{1}{2} \left ( \frac{\partial \hat{u}''_i}{\partial \xrog_j} + \frac{\partial \hat{u}''_j}{\partial \xrog_i} \right ) - \frac{1}{3} \frac{\partial \hat{u}''_k}{\partial \xrog_k} \delta_{ij} \right ] ,
\end{equation}
\begin{equation}
     \hat{E}_t = \hat{U}+ \hat{K}_t ,
\end{equation}
\begin{equation} 
    \hat{U} = C_v \hat{T} \qquad \hat{K}_t = \frac{1}{2} \hat{u}''_i \hat{u}''_i ,
\end{equation}
\begin{equation}
    \hat{P} = \hat{\rho} R \hat{T} ,
\end{equation}
\begin{equation}
    \hat{\kappa} = \frac{ \hat{\mu} C_p}{Pr} ,
\end{equation}
\begin{equation}
    \hat{\mu} = \mu_0 \left(\frac{ \hat{T}}{T_0}\right)^{n} ,
\end{equation}
\begin{equation}
\label{eq:frho_modframe}
\hat{f}^{(\rho)} = -2 \dot{L} \hat{\rho} ,
\end{equation}
\begin{equation}
\label{eq:frhou_modframe}
\hat{f}^{(u)}_i = -3 \dot{L} \hat{\rho} \hat{u}''_i ,
\end{equation}
\begin{equation}
\label{eq:fe_modframe}
\hat{f}^{(e)} = - \left [2 \hat{\rho} \hat{E}_t + \hat{\rho} \hat{u}''_i \hat{u}''_i + 3 \hat{P} \right ] \dot{L} .
\end{equation}

The last issue to be addressed is the time $\hat{t}$ that corresponds to $L=0$. Solving $\frac{dg}{dt} = L^{-1}$ leads to
\begin{equation}
g = -\frac{1}{2V_b} \ln( L).
\end{equation}
Since we evaluated the equations at time $t = g^{-1}(\hat{t})$, we have
\begin{equation}
\hat{t} = -\frac{1}{2V_b} \ln( L).
\end{equation}
Thus, $L=0$ corresponds to $\hat{t} \to \infty$. However, it is not expected that the simulation will need to proceed up to infinity, and that instead the viscous instability would kick in prior to this limit.

\section{Derivation of evolution equations for solenoidal and dilatational TKE}
\label{sec:sol_dil_evolution}
To derive the evolution equations for the solenoidal and dilatational TKEs given homogeneous turbulence with some applied mean-flow deformation, we follow the procedure of \cite{kida1990}. Thus, we introduce the variable $v_i = \sqrt{\rho} u''_i$. The derivation of the evolution equation for $v_i$ begins with
\begin{multline*}
\frac{\partial v_i}{\partial t} = \frac{\partial}{\partial t} \left ( \frac{\rho u''_i}{\sqrt{\rho}} \right) = \frac{1}{\sqrt{\rho}} \left ( \frac{\partial \rho u''_i}{\partial t} - \frac{u''_i}{2} \frac{\partial \rho}{\partial t} \right ) = \\
\frac{1}{\sqrt{\rho}} \left ( -\frac{\partial \rho u''_i}{\partial x_j} G_{jk}x_k - \frac{\partial \rho u''_i u''_j}{\partial x_j} + \frac{\partial \sigma_{ij}}{\partial x_j} + f_i^{(u)} + \frac{u_i}{2} \frac{\partial \rho}{\partial x_j} G_{jk}x_k + \frac{u_i}{2} \frac{\partial \rho u''_j}{\partial x_j} - \frac{u_i}{2} f^{(\rho)} \right ),
\end{multline*}
where we have used \cref{eq:rho_f_G,eq:rhou_f_G}. One can easily derive the relations
\begin{equation*}
-\frac{\partial \rho u''_i}{\partial x_j} G_{jk} x_k + \frac{u_i}{2} \frac{\partial \rho}{\partial x_j} G_{jk} x_k = -\sqrt{\rho} \frac{\partial v_i}{\partial x_j} G_{jk} x_k,
\end{equation*}
and
\begin{equation*}
- \frac{\partial \rho u''_i u''_j}{\partial x_j} +  \frac{u_i}{2} \frac{\partial \rho u''_j}{\partial x_j} = -\sqrt{\rho} \left ( \frac{\partial v_i u''_j}{\partial x_j} + \frac{ v_i}{2} \frac{\partial u''_j}{\partial x_j} \right),
\end{equation*}
to thus obtain
\begin{equation*}
\frac{\partial v_i}{\partial t} = -\frac{\partial v_i}{\partial x_j} G_{jk} x_k - \frac{\partial v_i u''_j}{\partial x_j} + \frac{v_i}{2} \frac{ \partial u''_j}{\partial x_j} + \frac{1}{\sqrt{\rho}} \frac{\partial \sigma_{ij}}{\partial x_j} + \frac{1}{\sqrt{\rho}} f_i^{(u)} - \frac{1}{\sqrt{\rho}} \frac{u''_i}{2} f^{(\rho)}.
\end{equation*}

The evolution equation for $v_i$ can now be multiplied by $v_i^{(\alpha)} = \sqrt{\rho} u''^{(\alpha)}_i$, where $\alpha = s,d$. This is then followed by the application of the averaging operator. Given the Helmholtz decomposition of any two fluctuating variables $F''_i = F''^{(s)}_i + F''^{(d)}_i$ and $G''_i = G''^{(s)}_i + G''^{(d)}_i$, the orthogonality relation $\langle F''^{(\alpha)}_i G''^{(\beta)}_i \rangle = 0$ for $\alpha,\beta = s,d$ and $\alpha \neq \beta$ is used to obtain
\begin{multline*}
\frac{1}{2} \frac{d}{d t} \lbra v_i^{(\alpha)} v_i^{(\alpha)} \rbra  = -\frac{1}{2} \frac{ \partial}{\partial x_j} \lbra v_i^{(\alpha)} v_i^{(\alpha)} \rbra  G_{jk} x_k \\
- \lbra \frac{\partial v_i u''_j}{\partial x_j} v_i^{(\alpha)} \rbra + \lbra \frac{v_i}{2} \frac{ \partial u''_j}{\partial x_j} v_i^{(\alpha)} \rbra + \lbra \frac{1}{\sqrt{\rho}} \frac{\partial \sigma_{ij}}{\partial x_j} v_i^{(\alpha)} \rbra + \lbra \frac{1}{\sqrt{\rho}} f_i^{(u)} v_i^{(\alpha)} \rbra - \lbra \frac{1}{\sqrt{\rho}} \frac{u''_i}{2} f^{(\rho)} v_i^{(\alpha)} \rbra.
\end{multline*}
The first term on the right-hand side above vanishes due to homogeneity. Plugging in for $v_i$ and using \cref{eq:frho,eq:frhou}, the above becomes 
\begin{multline*}
\frac{1}{2} \frac{d}{d t} \lbra \rho u''^{(\alpha)}_i u''^{(\alpha)}_i \rbra  = - \lbra \frac{\partial \sqrt{\rho} u''_i u''_j}{\partial x_j} \sqrt{\rho} u''^{(\alpha)}_i \rbra + \lbra \frac{\rho u''_i u''^{(\alpha)}_i}{2} \frac{ \partial u''_j}{\partial x_j} \rbra \\
+ \lbra  \frac{\partial \sigma_{ij}}{\partial x_j} u''^{(\alpha)}_i \rbra - \lbra \rho u''^{(\alpha)}_i u''^{(\alpha)}_j \rbra G_{ij} - \frac{1}{2} \lbra \rho u''^{(\alpha)}_i u''^{(\alpha)}_i \rbra G_{jj} .
\end{multline*}
Using either $\alpha = s$ or $d$, and assuming $G_{ij}$ is isotropic, we obtain the evolution equations for the solenoidal and dilatational kinetic energies, which we express as follows
\begin{align*}
\frac{d \langle \rho \rangle k^{(s)}}{d t} +\lbra \rho \rbra k^{(s)} G_{jj} &= IA^{(s)} +P^{(s)} - \langle \rho \rangle \epsilon^{(s)},  \\
\frac{d \langle \rho \rangle k^{(d)}}{d t} + \lbra \rho \rbra k^{(d)} G_{jj} &= IA^{(d)} +P^{(d)} - \langle \rho \rangle \epsilon^{(d)} + PD.
\end{align*}
In the above, $k^{(s)}$, $IA^{(s)}$, $P^{(s)}$, and $\epsilon^{(s)}$ are the solenoidal components of the turbulent kinetic energy, intermode advection, production,  and dissipation, respectively. $k^{(d)}$, $IA^{(d)}$, $P^{(d)}$, and $\epsilon^{(d)}$ are the dilatational components of the turbulent kinetic energy, intermode advection, production, and dissipation, respectively. $PD$ is the pressure dilatation. These quantities are defined as 
\begin{align}
k^{(\alpha)} &= \frac{1}{2}\ka ,\\
IA^{(\alpha)} & = - \left < \frac{\partial \sqrt{\rho} u''_i u''_j}{\partial x_j} \sqrt{\rho} u''^{(\alpha)}_i \right > + \left < \frac{\rho u''_i u''^{(\alpha)}_i}{2} d \right > ,\\
P^{(\alpha)} & = - \lbra \rho u''^{(\alpha)}_i u''^{(\alpha)}_j \rbra G_{ij} ,\\
\langle \rho \rangle \epsilon^{(s)} &= \left < \mu w_i w_i \right > ,\\
\langle \rho \rangle \epsilon^{(d)} &= \frac{4}{3} \left < \mu d^2 \right > ,\\
PD &= \left < P d \right > .
\end{align}
We note that
\begin{equation}
IA^{(s)} + IA^{(d)} = - \frac{\partial}{\partial x_j} \left ( \frac{1}{2} \left < \rho u''_i u''_i u''_j \right > \right ) = 0 ,
\end{equation} 
and thus, the advection terms represent an intermode transfer of energy between the solenoidal and dilatational components. Using the Favre-averaged conservation of mass equation, we finally express the evolution equation for the turbulent kinetic energies as  
\begin{align}
\lbra \rho \rbra \frac{d k^{(s)}}{d t} &= IA^{(s)} +P^{(s)} - \langle \rho \rangle \epsilon^{(s)} ,\\
\lbra \rho \rbra \frac{d k^{(d)}}{d t} &= IA^{(d)} +P^{(d)} - \langle \rho \rangle \epsilon^{(d)} + PD.
\end{align}

\section{Proof of time invariance for the energy ratio $\left ( \widetilde{U} + k \right )/ \widetilde{U}^{(a)}$}
\label{sec:cons_energy_ratio}
The chain rule applied to the time derivative of the energy ratio gives
\begin{equation}
\label{eq:ratio_1}
\frac{d}{dt} \left ( \frac{ \widetilde{U} + k }{ \widetilde{U}^{(a)}} \right ) = \frac{d}{dt} \left ( \widetilde{U} + k \right ) \frac{1}{\widetilde{U}^{(a)}} + \left ( \widetilde{U} + k \right ) \frac{d}{dt} \left ( \frac{1}{\widetilde{U}^{(a)}} \right ).
\end{equation}
Given the definition of the adiabatic internal energy $\widetilde{U}^{(a)} = \widetilde{U}_0 L^{-2}$, we have
\begin{equation}
\label{eq:ratio_2}
\frac{d}{dt} \left ( \frac{1}{\widetilde{U}^{(a)}} \right ) = \frac{2 L \dot{L}}{\widetilde{U}_0} .
\end{equation}
Using \cref{eq:ks_evolution,eq:kd_evolution,eq:ie_evolution}, one obtains
\begin{equation}
\frac{d}{dt} \left ( \widetilde{U} + k \right ) = \frac{MW + P}{\lbra \rho \rbra} ,
\end{equation}
where $P$ is the total production $P^{(s)} + P^{(d)}$. Given the deformation tensor $G_{ij}$ used for isotropic compressions, $MW = -3 \lbra P \rbra \dot{L}/L$ and $P = -2 \lbra \rho \rbra k \dot{L} / L$. Using the equation of state $\lbra P \rbra = \lbra \rho \rbra R \widetilde{T}$, the definition of the internal energy $\widetilde{U} = C_v \widetilde{T}$, and the specific heat ratio $\gamma = 5/3$, we have
\begin{equation}
\label{eq:ratio_3}
\frac{d}{dt} \left ( \widetilde{U} + k \right ) = -2 \widetilde{U} \frac{\dot{L} }{L} - 2 k \frac{\dot{L}}{L} .
\end{equation}
Thus, using \cref{eq:ratio_2,eq:ratio_3} in \cref{eq:ratio_1}, we can show that
\begin{equation}
\frac{d}{dt} \left ( \frac{ \widetilde{U} + k }{ \widetilde{U}^{(a)}} \right ) = -2 \widetilde{U} \frac{\dot{L}{L}}{\widetilde{U}_0} - 2k \frac{\dot{L} L}{\widetilde{U}_0} + 2 \widetilde{U} \frac{ L \dot{L}}{\widetilde{U}_0} + 2 k \frac{ L \dot{L}}{\widetilde{U}_0} = 0 .
\end{equation}

\bibliographystyle{unsrt}
\bibliography{library}

\end{document}